\documentclass[a4paper,fleqn,usenatbib]{mnras}

\usepackage{mathptmx}

\usepackage[T1]{fontenc}
\usepackage{ae,aecompl}

\usepackage{graphicx}	
\usepackage{amsmath}	
\usepackage{amssymb}	

\usepackage[svgnames]{xcolor}

\usepackage[normalem]{ulem}

\usepackage{soul}
\definecolor{chartreuse}{rgb}{0.87, 1.0, 0.0}
\sethlcolor{chartreuse}

\title[The life cycles of Be viscous decretion discs]{The life cycles of Be viscous decretion discs:\\The case of $\omega$ CMa}

\author[M. R. Ghoreyshi et al.]{
\parbox{\textwidth}{M. R. Ghoreyshi,$^{1,2}$\thanks{Email: mohammad@usp.br}
A. C. Carciofi,$^{1}$
L. R. R\'{\i}mulo,$^{1}$
R. G. Vieira,$^{1}$
D. M. Faes,$^{1}$
D. Baade,$^{3}$
J. E. Bjorkman,$^{4}$
S. Otero$^{5}$
and Th. Rivinius$^{6}$} \vspace{0.4cm}\\
\parbox{\textwidth}{
$^{1}$Instituto de Astronomia, Geof\'{\i}sica e Ci\^{e}ncias Atmosf\'{e}ricas, Universidade de S\~{a}o Paulo, Rua do Mat\~{a}o 1226, S\~{a}o Paulo, SP 05508-900, Brazil\\
$^{2}$Research Institute for Astronomy and Astrophysics of Maragha (RIAAM), Maragha, P.O. Box: 55134-441, Iran\\
$^{3}$European Organisation for Astronomical Research in the Southern Hemisphere (ESO), Karl-Schwarzschild-Str. 2, 85748 Garching b. M\"{u}nchen, Germany\\
$^{4}$Department of Physics  Astronomy, University of Toledo, MS111 2801 West Bancroft Street, Toledo, OH 43606, USA\\
$^{5}$American Association of Variable Star Observers (AAVSO), Cambdrige, MA, USA\\
$^{6}$European Organisation for Astronomical Research in the Southern Hemisphere (ESO), Casilla 19001, Santiago 19, Chile\\
}}

\date{Accepted XXX. Received YYY; in original form ZZZ}

\pubyear{2017}

\begin{document}
\label{firstpage}
\pagerange{\pageref{firstpage}--\pageref{lastpage}}
\maketitle

\begin{abstract}
We analyzed $V$-band photometry of the Be star $\omega$ CMa, obtained during the last four decades, during which the star went through four complete cycles of disc formation and dissipation. The data were simulated by hydrodynamic models based on a time-dependent implementation of the viscous decretion disc (VDD) paradigm, in which a disc around a fast-spinning Be star is formed by material ejected by the star and driven to progressively larger orbits by means of viscous torques. Our simulations offer a good description of the photometric variability during phases of disc formation and dissipation, which suggests that the VDD model adequately describes the structural evolution of the disc. Furthermore, our analysis allowed us to determine the viscosity parameter $\alpha$, as well as the net mass and angular momentum (AM) loss rates. We find that $\alpha$ is variable, ranging from 0.1 to 1.0, not only from cycle to cycle but also within a given cycle. Additionally, build-up phases usually have larger values of $\alpha$ than the dissipation phases. Furthermore, during dissipation the outward AM flux is not necessarily zero, meaning that $\omega$ CMa does not experience a true quiescence but, instead, switches between a high to a low AM loss rate during which the disc quickly assumes an overall lower density but never zero. We confront the average AM loss rate with predictions from stellar evolution models for fast-rotating stars, and find that our measurements are smaller by more than one order of magnitude.
\end{abstract}

\begin{keywords}
Techniques: photometric -- Stars: emission-line, Be -- Stars: mass-loss -- Stars: individual: $\omega$ CMa
\end{keywords}



\section {Introduction}
\label{introduction}

During the last decade, the equatorial, dust-free, Keplerian gaseous discs present around some rapidly rotating B stars were resolved by long-baseline optical and near-infrared (NIR) interferometry. Also, these discs reveal themselves by H\,{\sc i} emission lines superimposed to the optical and IR spectrum of the star. Therefore, owing to the presence of emission lines, this subclass of non-supergiant, main-sequence B stars are specifically called Classical Be stars. These stars were the subject of a recent review paper by \cite{rivinius2013a}.

Conceptually, a Be disc is formed in two steps: the disc is fed by material ejected from the star with enough angular momentum (hereafter, AM) to reach an orbit around the star; subsequently the gas diffuses outwards.
Recently, \cite{baade2016} shed some light on the disc feeding mechanism: the authors suggest that a combination of frequencies of different pulsation modes may provide the needed kinetic energy and AM. After material is ejected, it slowly diffuses outwards by means of viscous torques. Thus, the discs around Be stars are likely described by the same mechanism found in the accretion discs of young stellar objects, namely viscous transport of material and AM. However, while in accretion discs the direction of mass flow is inwards, in Be stars it can alternate between outflow (decretion, when the disc is being actively fed by the central star) and inflow (when the disc reaccretes a large fraction of the material back to the star after the feeding mechanism is turned off), and for this reason the discs are referred to by the neologism ``decretion discs''. In practice, Be discs can be quite complicated, because different parts of the disc (e.g., inner vs. outer) may display different behaviors (accretion vs. decretion, see, e.g., \citealt{haubois2012})

Accretion discs are seen in active galactic nuclei (AGNs) and protostellar clouds. The accretion process also occurs in a binary system when one of the companions fills its Roche lobe and spills material onto the secondary, which can be a black hole, a neutron star, a white dwarf or a dwarf star. Such a wide variety of systems, in which viscosity is of paramount importance, provides the motivation to investigate viscosity and measure its value; here, Be stars are ideal testbeds because their discs are dust-free and possess a considerably simpler structure than the aforementioned examples. Also, Be discs are relatively small (with typical sizes of a few tens of stellar radii), which implies that they have much shorter evolutionary timescales (weeks to months) than, say, AGN or protostellar discs. Furthermore, thanks to their specific properties, the study of Be stars provides a unique opportunity to probe and understand several important branches of astrophysics, e.g., asymmetric mass-loss processes, evolution of fast-rotating stars, asteroseismology, and, what is the most relevant for this paper, astrophysical discs.

One open issue for such systems is the origin of viscosity: so far no satisfactory theory of viscosity exists for the circumstellar medium and for this reason the $\alpha$ prescription for viscosity is commonly assumed \citep{shakura1973}. The $\alpha$ parameter links the scale of the turbulence to the (vertical) scale of the disc by the following formula
\begin{equation}
\label{alpha_definition}
\nu = \frac{2}{3}\alpha c_{s}H,
\end{equation}
where $\nu$ is the viscosity, $c_{s}$ is the isothermal sound speed and $H$ is the disc scaleheight. Measuring the value of $\alpha$ as directly as possible will help astronomers to have a better understanding of the physics of viscosity and also of viscous discs, and this is one of the main motivations for this work. 

Viscous transport of material and AM forms the basis of the Viscous Decretion Disc model \citep{lee1991}. The model has been successfully applied to study individual stars (for instance, \citealt{carciofi2006b}; \citealt{jones2008}; \citealt{carciofi2009}; \citealt{klement2015}, \citeyear{klement2017}) and samples of Be stars (for instance, \citealt{silaj2010}; \citealt{touhami2011}; \citealt{vieira2017}; \citealt{rimulo2018}). The main previous results are summarized by \cite{rivinius2013a}.

The first effort to understand the dynamical evolution of VDDs around isolated Be stars was done by \cite{jones2008}. This was later followed by a systematic study by \cite{haubois2012} who coupled the {\tt SINGLEBE} \citep{okazaki2007} and {\tt HDUST} (\citealt{carciofi2006a}, \citeyear{carciofi2008}) codes to study the theoretical effects of time variable mass loss rates on the structure of the disc and its consequences on the observed photometry. An important conclusion of their work was that time-dependent VDD calculations can explain well the observed temporal photometric phenomenology of Be stars such as loops in the color-magnitude diagram, believed to track the process of the disc formation, during which the stars become redder and brighter, and dissipation, as the stars move back to their intrinsic colors and brightness. However, the tracks in the color-magnitude diagram are often complex and dependent on the inclination angle, as discussed by \cite{haubois2012}

This conclusion gained further support when \cite{carciofi2012} (hereafter C12) successfully modeled the light curve of a Be star. Their fit of the dissipation phase following the third outburst of $\omega$ CMa (see Sect.~\ref{observations}) allowed the authors to determine $\alpha=1.0\pm 0.2$.

$\omega$ CMa is one of the brightest Be stars ($m_\mathrm{v} \approx$ 3.6 to 4.2) in the sky and it has been a common target of observers. Therefore, there exists a rich dataset that has been investigated by several authors. In this paper we build upon C12's work by modeling the full light curve of $\omega$ CMa since late 1981 to late 2015. In the period of 34 years, the star went through four complete cycles of disc formation and dissipation. Our main goals are i) to verify that the VDD model can provide an adequate description of the photometric behavior of $\omega$ CMa during both the phases of disc formation and dissipation; ii) to determine the mass and AM loss rates associated with the disc formation phases; and, iii) to measure the viscosity parameter for each phase (outburst vs. quiescence) at each individual cycle. 

In Sect.~\ref{observations} we present the available $V$-band data of $\omega$ CMa. In Sect.~\ref{theory} the theoretical framework adopted for the VDD is introduced. Sect.~\ref{cycle_lengths} presents a model-independent analysis of the light curve which was used as input for the VDD modeling presented in Sect.~\ref{vdd_model}. In Sect.~\ref{discussion} the results are discussed and our main conclusions are summarized in Sect.~\ref{conclusions}.


\begin{figure*}
\centering
\includegraphics[width=1.0\linewidth, height=0.45\linewidth]{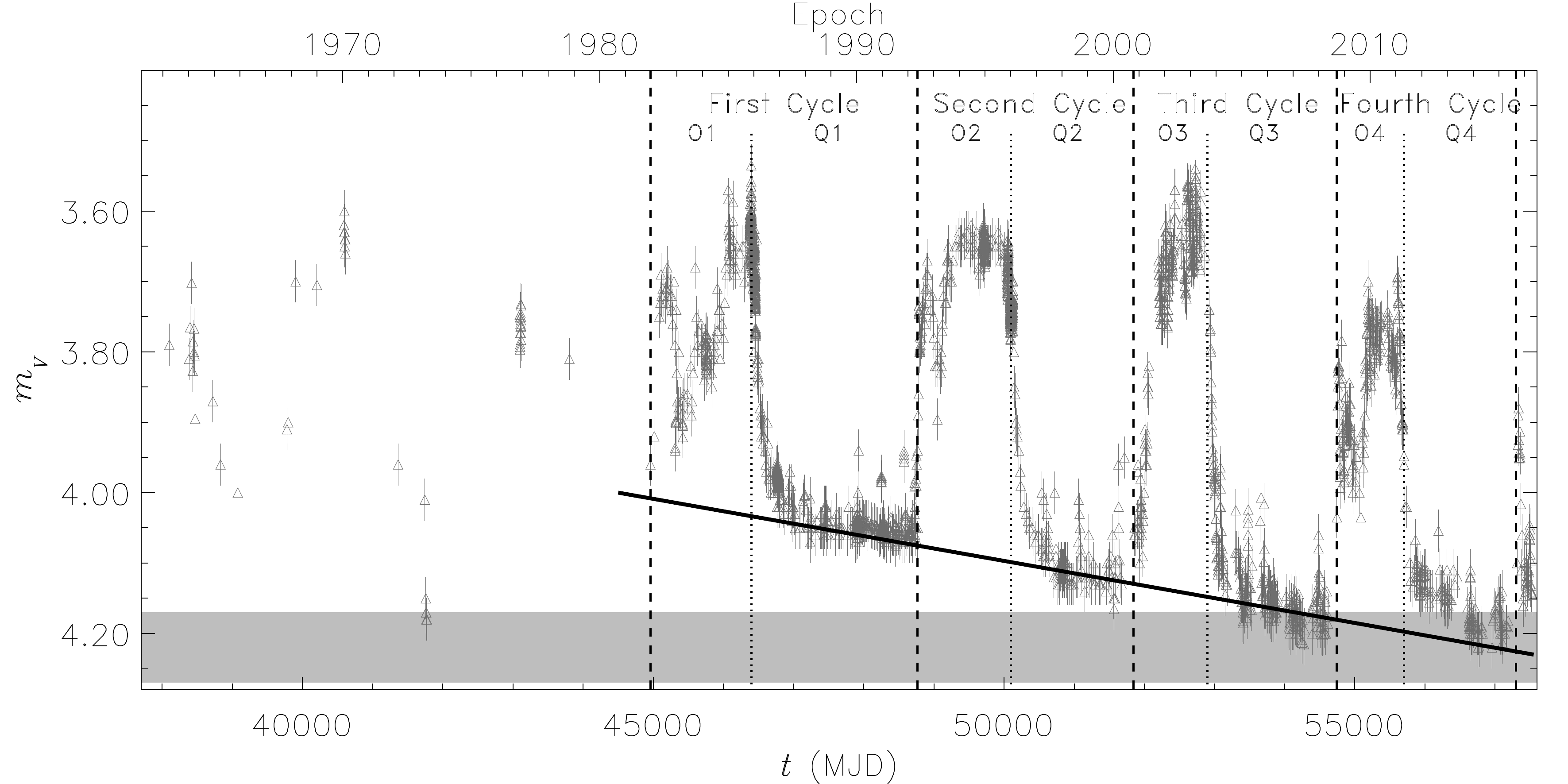}
\caption{
$V$-band light curve of $\omega$ CMa (grey triangles), showing four full cycles, as indicated. The thick solid line is a linear fit of the last 100\,d of each cycle, indicating that the lowest brightness of each subsequent cycle is fainter than the previous one. O$i$ and Q$i$ stand for outburst and quiescence phases, respectively, where $i$ is the cycle number. The light curve is a collection of observations from the following sources: \citet{baade1982}, \citet{stagg1987}, \citet{dachs1988}, \citet{edalati1989}, photoelectric observations obtained in the Long-Term Photometry of Variables (\citealt{manfroid1991}, \citeyear{manfroid1995}; \citealt{sterken1993}), \citet{mennickent1994}, Hipparcos (\citealt{perryman1997}), \citet{stefl2000}, and the visual observations by Otero \citep{stefl2003a}. The horizontal grey band represents the estimated intrinsic visual magnitude of the central star of $\omega$ CMa. The vertical dashed and dotted lines indicate the transitions between quiescence and outburst (and vice-versa).}
\label{complete_lightcurve}
\end{figure*}



\section {Observations}
\label{observations}

$V$-band photometric data of $\omega$ (28) CMa (HD 56139, HR2749; B2 IV-Ve) are available from many sources starting in 1963. Unfortunately, the data from 1963 to 1982 are too sparse, so we focus on the data from 1982 onwards. The majority of the data correspond to visual observations using a modified version of the Argelander method \citep{hirshfeld1985}, carried out from 1997 to the present. To increase the accuracy for estimating the small magnitude variations of $\omega$ CMa, a grid of standard stars (Table 2 in \citealt{stefl2003a}) was used to determine the visual magnitude. Thus, the availability of nearby comparison stars governs the accuracy that is typically better than 0$\fm$05. Another important part of the data were observed by Edalati and his professional and amateur collaborators during a survey of variable stars \citep{edalati1989}. The full assembled light curve is shown in Fig.~\ref{complete_lightcurve}. The sources for the data are listed in the caption of the figure.


\begin{figure*}
\centering
\includegraphics[width=1.0\linewidth, height=0.45\linewidth]{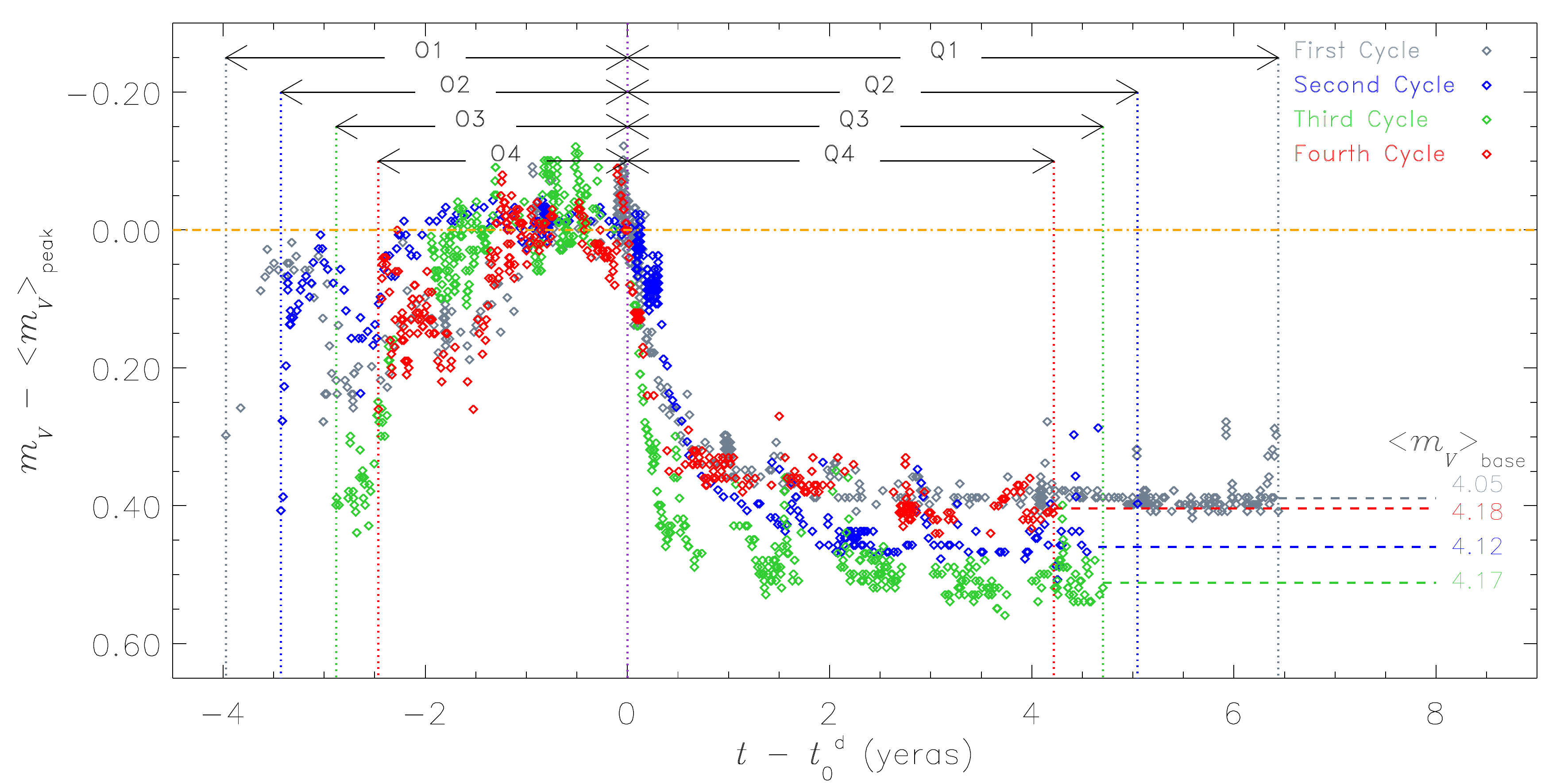}
\caption{$V$-band photometry of $\omega$ CMa. The data in different cycles are displaced horizontally in time so that the beginnings of all four dissipation phases are roughly aligned with the onset of the disc dissipation (vertical violet dotted line), and vertically so that the average magnitude of the last year of outburst is zero ($\langle m_\mathrm{v}\rangle_\mathrm{peak}$=0). Also, $\langle m_\mathrm{v}\rangle_\mathrm{base}$ is the average magnitude for the last two years of each dissipation phase. Each cycle is shown by an individual color as indicated. The epoch of onset of each dissipation phase ($t_0^{\rm d}$) was determined in Sect.~\ref{cycle_lengths}, and is listed in Table~\ref{results}.}
\label{overlapped}
\end{figure*}


Since 1982, $\omega$ CMa exhibited four quasi-regular cycles with lengths varying between $\sim$7.0 to $\sim$10.5 years. Each cycle includes two main parts: 1) An outburst phase characterized by a brightening of 0$\fm$3 to 0$\fm$5 during $\sim$ 2.5--4.0 years and a reddening of $B-V$ $\approx$ -- 0$\fm$15 to $\approx$ -- 0$\fm$10 \citep{stefl2003a}, and 2) a quiescence phase that lasts about $\sim$4.5--6.5 years. Throughout this text we refer to the cycles by C$i$ and to the phases by O$i$ and Q$i$ for outburst and quiescence, respectively, where $i$ is the cycle number.

An inspection by eye of the light curve of $\omega$ CMa reveals some noticeable features. First of all, one can easily see a long-term secular decline in the brightness of the system in successive dissipation phases (the solid line in Fig.~\ref{complete_lightcurve}). The difference between the minimum brightnesses of the first cycle and the fourth one is about 0$\fm$12. Moreover, outbursts and quiescence of different cycles clearly have different lengths and amplitudes. The main outburst phases are not smooth, but instead develop in steps, and several individual short-term (typically 30-day long) outbursts can be identified. The quiescence phases, however, are much smoother, and some short-term small-scale outbursts are also observed.

In order to better identify the cycle-to-cycle variations, in  Fig.~\ref{overlapped} all four cycles are over plotted, shifted both in time and in magnitude. Figure~\ref{overlapped} reveals that over the last 34 years, i) the cycle length is decreasing (from about 10.5 years for the first cycle to about 7.0 years for the fourth cycle); ii) The duration of the outbursts and quiescences are correlated: the longer the outburst the longer the subsequent quiescence phase; iii) The disc dissipation rate is faster in the last two cycles than in the first two.


\section {Theoretical Considerations}
\label{theory}


\subsection {Model description}
\label{model}

\cite{maintz2003} presented a detailed analysis of the line profile variability of $\omega$ CMa, which is a single-mode nonradial pulsator \citep{stefl2003b}. They conclude that the star is viewed nearly pole-on ($i=15^{\circ}$). Their determination for the stellar parameters will be used in this work, and are listed in Table~\ref{stellar_parameter}. The star's apparent magnitude, $m^{*}_\mathrm{v}$, was calculated with {\tt HDUST} based on the adopted stellar parameters, assuming there is no disc around the central star. The error in $m^{*}_\mathrm{v}$ was estimated by considering a $\pm 1 \sigma$ uncertainty in the distance.


\begin{table}
\begin{center}
\caption{The stellar parameters of $\omega$ CMa.}
\begin{tabular}{@{}cccccccccc}
\hline
\hline
Parameter & & & & Value & reference \\
\hline
$L$ & & & & 5224 L$_{\odot}$ & \citealt{maintz2003} \\
$T_\mathrm{pole}$ & & & & 22000 K & \citealt{maintz2003} \\
$R_\mathrm{pole}$ & & & & 6.0 R$_{\odot}$ & \citealt{maintz2003} \\
log g$_\mathrm{pole}$ & & & & 3.84 & \citealt{maintz2003} \\
$M$ & & & & 9.0 M$_{\odot}$ & \citealt{maintz2003} \\
$V_\mathrm{rot}$ & & & & 350 km s$^{-1}$ & \citealt{maintz2003} \\
$V_\mathrm{crit}$ & & & & 436 km s$^{-1}$& \citealt{maintz2003} \\
$R_\mathrm{eq}$ & & & & 7.5 R$_{\odot}$ & \citealt{maintz2003} \\
$i$ & & & & 15$^{\circ}$ & \citealt{maintz2003} \\
distance & & & & 279 $\pm$ 14 pc & \citealt{perryman1997}\\
$m^{*}_\mathrm{v}$ & & & & 4.22 $\pm$ 0.05 & This work \\

\hline
\end{tabular}
\label{stellar_parameter}
\end{center}
\end{table}


For the physical modeling of the light curve of $\omega$ CMa we used the radiative transfer code {\tt HDUST} along with the time-dependent 1-D $\alpha$-disc code {\tt SINGLEBE} which provides the disc surface density and the mass flux as a function of distance from the star at each selected time. This, in turn, is used as input for {\tt HDUST} to calculate the emergent spectrum as a function of time. The modeling procedure follows C12, \cite{haubois2012} and \cite{rimulo2018}, and is only summarized here.

For this paper the relevant quantities are the mass flux, $\dot{M}_\mathrm{disc}$, and the AM flux, $\dot{J}_\mathrm{disc}$. Assuming a Keplerian motion for the gas \citep[e.g.,][]{pringle1981, meilland2007}, the mass flux is related to the disc surface density, $\Sigma(r, t)$, by the mass conservation relation
\begin{equation}
\label{mdot_disc}
\dot{M}_\mathrm{disc}(r,t) = 2\pi r \Sigma(r, t) v_r = -4\pi \left(\frac{r^3}{GM_*}\right)^\frac{1}{2}\frac{1}{r}\frac{\partial}{\partial r}\left(\alpha c_s^2 r^2\Sigma(r, t)\right),
\end{equation}
where $v_r$ is the radial speed, $G$ is the universal gravitational constant, $M_*$ is the stellar mass, $\alpha$ is the viscosity parameter, $c_\mathrm{s}$ is the isothermal sound speed, given by $c_s^2=kT_\mathrm{disc}/\mu m_H$, and $r$ is the distance from the star. The AM flux, according to the AM conservation relation, is given by
\begin{equation}
\label{jdot_disc}
\dot{J}_\mathrm{disc}(r,t) = 2\pi r \Sigma(r, t) v_r \left(GM_*r\right)^\frac{1}{2} + 2\pi \alpha c_s^2 r^2\Sigma(r, t)\,,
\end{equation}
where the first term is the AM flux that is carried with the radial motion of the gas and the second term is the AM flux due to the torque generated by the viscous force.

The temporal evolution of the surface density is given by the following diffusion-like equation, which holds in the thin disc approximation ($c_\mathrm{s}^2\ll GM_*/r$):

\begin{equation}
\label{sigmadot}
\frac{\partial{\Sigma}}{\partial{t}}=\frac{2}{\bar{r}} \frac{\partial}{\partial{\bar{r}}} \left\{ \bar{r}^\frac{1}{2}\frac{\partial{}}{\partial{\bar{r}}}\left[\frac{\alpha c_\mathrm{s}^2}{(GM_*R_\mathrm{eq})^\frac{1}{2}} \bar{r}^2 \Sigma(r, t)\right]\right\}+S_\Sigma,
\end{equation}
where $R_\mathrm{eq}$ is the stellar equatorial radius, $\bar{r}$ is the normalized radius ($r/R_\mathrm{eq}$) and $S_\Sigma$ represents the mass injection rate per unit area from the star into the disc. 

In this work, it was assumed that the star injects mass in a Keplerian orbit at a given rate, $\dot{M}_\mathrm{inj}$, at a radius very close to the surface of the star, $R_\mathrm{inj}$ = 1.02 $R_\mathrm{eq}$, at the equatorial plane (see below). This assumption may be too simplistic for some Be stars, because there is observational evidence pointing to asymmetric mass loss (e.g., the short-term $V/R$ variations of $\eta$ Cen, \citealt{rivinius1997}) or matter being ejected at higher latitudes (e.g., \citealt{stefl2003a}). However, this is not an issue for this work, since after a few orbital periods, the matter loses memory of the injection process owing to viscous diffusion and orbital phase mixing. Hence, $S_\Sigma$ is assumed to be given by 
\begin{equation}
\label{s_sigma}
S_\Sigma=\frac{\dot{M}_\mathrm{inj}}{2\pi R_\mathrm{eq}^2}\frac{\delta(\bar{r}-\bar{r}_\mathrm{inj})}{\bar{r}},
\end{equation}
where $\bar{r}_\mathrm{inj}=R_\mathrm{inj}/R_\mathrm{eq}$.

Finally, we assume torque-free boundary conditions at the stellar surface ($r=R_\mathrm{eq}$) and at a very distant outer radius ($r=R_\mathrm{out}$). As the second term in Eq.~(\ref{jdot_disc}) shows, this is accomplished by setting $\Sigma=0$ at those boundaries. The outer boundary, in particular, could represent the limiting radius of the disc due to a binary companion (e.g., \citealt{okazaki2002}) or due to the photoevaporation of the disc (e.g., \citealt{okazaki2001}). It is important to mention that we are only concerned with the very inner part of the disc, as the visual-flux excess originates in a very small disc volume close to the star (up to a few stellar radii, depending on the density scale, \citealt{carciofi2011}, \citealt{vieira2015}).

The conservation of AM through the system composed by the star, the disc and the outside medium can be written as follows
\begin{equation}
\label{jdot_star}
\dot{J}_*(t)+\frac{\mathrm{d}\phantom{t}}{\mathrm{d}t}\int_{R_\mathrm{eq}}^{R_\mathrm{out}}(GM_*r)^\frac{1}{2}\Sigma(r,t) 2\pi r\mathrm{d}r+\dot{J}_\mathrm{disc}(R_\mathrm{out},t)=0\,,
\end{equation}
where the first, second and third terms are the variation rates of the AM in the star, the disc and the outside medium, respectively. The second and third terms are obtained using the solution $\Sigma(r,t)$ of Eq.~(\ref{sigmadot}), which allows us to obtain the AM lost by the star (the first term).

In a steady-state decretion disc, AM is injected at a constant rate $(GM_* R_\mathrm{inj})^\frac{1}{2}\dot{M}_\mathrm{inj}$ at the radius $R_\mathrm{inj}$. This AM is divided into a constant AM flux inwards, for $r<R_\mathrm{inj}$, and a constant AM flux outwards, for $r>R_\mathrm{inj}$. Since in steady-state the AM in the disc (second term of Eq.~\ref{jdot_star}) is constant in time, the constant flux outwards in the region $r>R_\mathrm{inj}$ is the AM loss rate of the star. It is given by \cite{rimulo2018}
\begin{equation}
\label{jdot_steady}
-\dot{J}_{*,\mathrm{std}}=\Lambda \left(GM_*R_\mathrm{eq}\right)^\frac{1}{2}\dot{M}_\mathrm{inj}\left(\bar{r}_\mathrm{inj}^\frac{1}{2}-1\right)\,,
\end{equation}
where $\Lambda=1/(1-\bar{r}_\mathrm{out}^{-\frac{1}{2}})$, with $\bar{r}_\mathrm{out}=R_\mathrm{out}/R_\mathrm{eq}$, is a number usually just slightly larger than 1, as $R_\mathrm{out}\gg R_\mathrm{eq}$. 
The steady-state AM loss rate, therefore, depends very little (through the factor $\Lambda$) on the outer radius of the disc, whose value is poorly known for a few Be stars, and completely unknown for most \citep{klement2017}. In the simulations presented in this work, we have set $\bar{r}_\mathrm{out}=1000$.

The steady-state surface density for radii larger than $R_\mathrm{inj}$ is given by 
\begin{equation}
\label{sigma_steady}
\Sigma_\mathrm{std}(r)=\frac{-\dot{J}_{*,\mathrm{std}}}{2\pi\alpha c_s^2}\frac{1}{r^2}\left(1-\frac{r^\frac{1}{2}}{R_\mathrm{out}^\frac{1}{2}}\right)\,,\,R_\mathrm{inj}\leq r \leq R_\mathrm{out}\,,
\end{equation}
from which we see that the disc density, a physical quantity that can be estimated, for instance, from SED modeling, scales with $-\dot{J}_{*,\mathrm{std}}/\alpha$. Note that the $\Sigma_\mathrm{std}$ and $\dot{J}_{*,\mathrm{std}}$ are just another way of expressing the rate of mass and angular momentum injection into the disc and the ``std'' subscript is used just to stress that these were defined in the steady-state limit.

The dynamical evolution of the disc surface density (Eq.~\ref{sigmadot}) is driven by the time-variable source term of Eq.~(\ref{s_sigma}). It has been verified \citep{rimulo2018} that the solution of Eq.~(\ref{sigmadot}) for $r > \bar{r}_\mathrm{inj}$ is negligibly affected by variations of both $\dot{M}_\mathrm{inj}(t)$ and $\bar{r}_\mathrm{inj}$ as long as the product $\dot{M}_\mathrm{inj}(t)(\bar{r}_\mathrm{inj}^\frac{1}{2}-1)$ is kept unchanged. Since the radius $\bar{r}_\mathrm{inj}$ is a quite unknown quantity, it follows that the mass injection into the disc  is best described by the quantity defined in Eq.~(\ref{jdot_steady}), which roughly represents the net AM injected into the disc. In this work, we have arbitrarily set $\bar{r}_\mathrm{inj}=1.02$.


At outburst, $-\dot{J}_{*,\mathrm{std}}$ is non-zero, which results in an outflowing ($v_{r}>0$, $\dot{M}_\mathrm{disc} > 0$) decretion disc. During this phase, the disc is slowly built inside-out \citep{haubois2012}. Quiescence is usually understood as a phase where $-\dot{J}_{*,\mathrm{std}}=0$ (however, see Sect.~\ref{vdd_model}). At this phase, the inner disc passively diffuses inward ($v_{r}<0$, $\dot{M}_\mathrm{disc} < 0$) while the outer disc remains outflowing (see Fig. 10 of \citealt{haubois2012}). 

From Eq.~(\ref{sigmadot}), the rate at which the surface density changes depends not only on the viscosity parameter $\alpha$, but also on the stellar radius (which sets the size of the region over which diffusion happens), and the isothermal sound speed, which was estimated from the approximate relation $T_\mathrm{disc}\approx 0.6T_\mathrm{pole}$ \citep{carciofi2006a}. As $R_\mathrm{eq}$ and $T_\mathrm{pole}$ are well-known for $\omega$ CMa (see Table \ref{stellar_parameter}), it follows that the observed changes in the light curve, being the result of variations of the disc density, are controlled only by $-\dot{J}_{*,\mathrm{std}}$ and $\alpha$. This realization forms the basis of the present analysis: $-\dot{J}_{*,\mathrm{std}}$ and $\alpha$ are allowed to vary both in time and magnitude to reproduce the observed light curve. A byproduct of the hydrodynamic simulations is $\dot{M}_\mathrm{disc}$ as a function of time and distance to the star, from which we can determine the amount of mass and AM effectively lost through the disc.

The relation between the surface density, $\Sigma$, and the volume density, $\rho$, is given by
\begin{equation}
\rho(r,z,t)=\frac{\Sigma(r,t)}{(2\pi)^\frac{1}{2}H}e^{-\frac{z^2}{2H^2}}\,,
\label{rho}
\end{equation}
where $H(r)=c_s/(GM_*/r^3)^\frac{1}{2}$ is the disc hydrostatically supported scaleheight.


In this paper, Eq.~(\ref{sigmadot}) is solved by the {\tt SINGLEBE} code of \citet{okazaki2007} using the source term defined by Eq.~(\ref{s_sigma}) and (\ref{jdot_steady}). Given a prescription for $\dot{J}_{*,\mathrm{std}}$, which can vary in time at will, {\tt SINGLEBE} computes the corresponding temporal evolution of $\Sigma(r,t)$.


\subsection {Mass reservoir effect}
\label{mass_reservoir}

As previously mentioned, only the inner parts of the disc contribute significantly to the visible excess emission observed in Be stars. From this it follows that the visible excess is not a good indicator of the total disc mass. This is easily understood if one considers the case of a disc fed at a constant rate for a very long time. Initially the brightness grows very fast, as a result of the increase of the density of the inner disc, but after a few weeks/months the light curve flattens out as the inner disc reaches a near steady-state configuration. The outer parts, however, continue to grow in mass without any effects on the visible light curve. To observationally probe these parts, one would have to follow up the light curve in the IR or radio wavelengths (\citealt{vieira2015}, \citealt{panoglou2016}).

When the disc feeding eventually ceases, the viscous forces that couple all the matter in the disc lead to the reaccretion of part of the disc back onto the star. In that case, discs that are very similar in their inner regions (having, therefore, similar visual excesses) but have different masses (as a result of a different previous evolution) will dissipate at different rates, owing to the fact that the inner region is viscously coupled to the outer parts. More specifically, a more massive disc can supply the inner regions with mass for a longer time than a less massive disc. As a result, the dissipation of more massive discs appears slower than the dissipation of less massive ones.

\begin{figure}
\centering
{\includegraphics[width=1.0\linewidth]{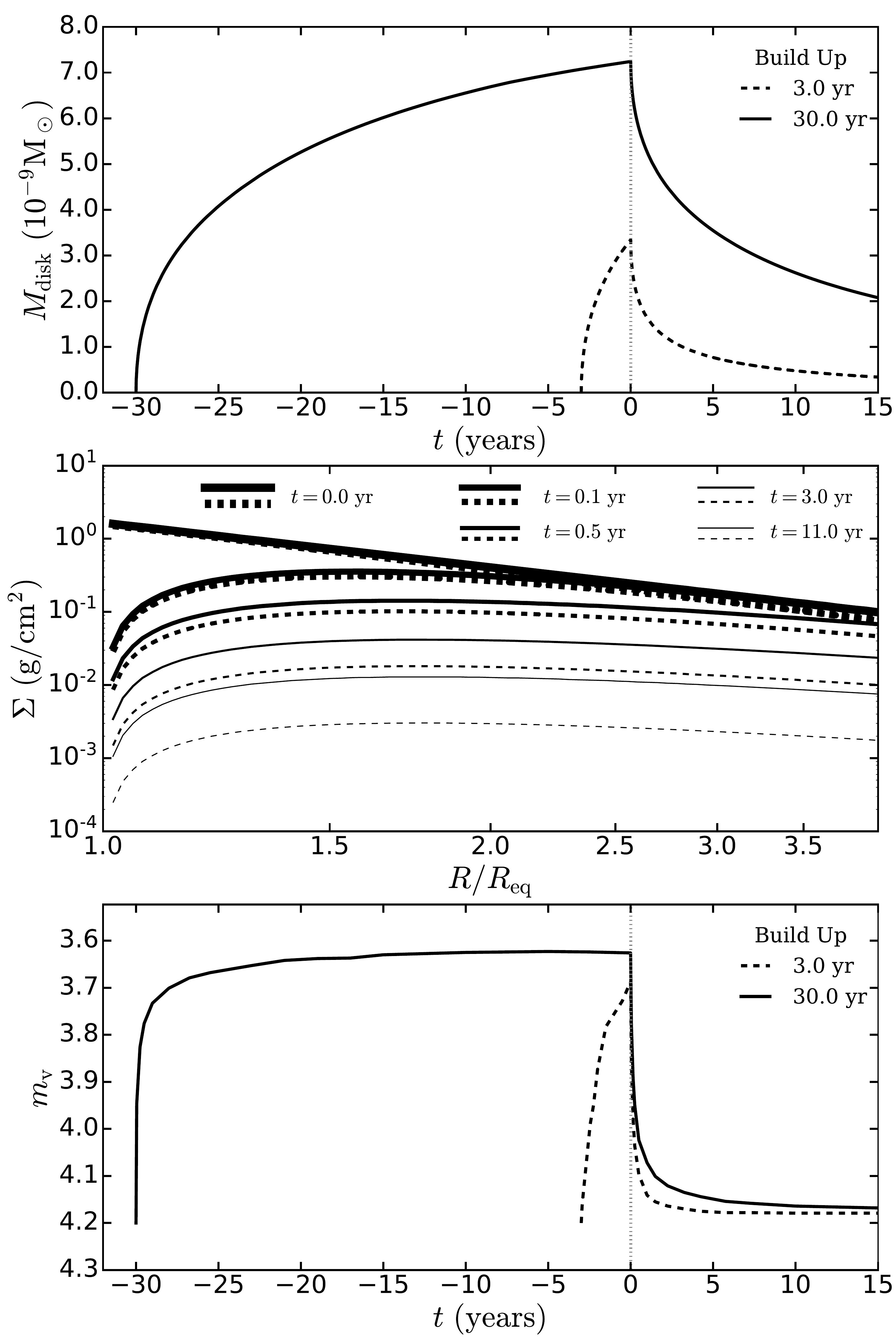}}
\caption{
Theoretical comparison between two events of discs formation and dissipation. In one event (solid lines) the disc was fed at a constant rate for 30.0\,yr. The dashed lines show another model for a much shorter build-up time (3.0\,yr). To facilitate comparison, we assumed $t=0$ as the onset of disc dissipation for both models.
 {\it Top}: The total disc mass of the two models as a function of time.
 {\it Middle}: Disc surface density as a function of distance from the star. The two lines trace the disc build-up for different times (as indicated) and the subsequent dissipation.  
 {\it Bottom}: The corresponding lightcurves for both models.
}
\label{disc_mass}
\end{figure}

In Fig.~\ref{disc_mass}, we compare two disc models that have been fed at the same rate but for different time spans ($3.0~\mathrm{yr}$ and $30.0~\mathrm{yr}$). In this comparison, we arbitrarily fixed the time $t=0$ as the end of the build-up phase. In both models, $-\dot{J}_{*,\mathrm{std}}=2.68\times 10^{36}\,\mathrm{g\,cm^2\,s^{-2}}$ and $\alpha=0.5$. The inner regions of both disc models reach a near-steady surface density by the time of the end of the smallest build-up time ($3.0~\mathrm{yr}$), thus, the surface density for the solid and dashed lines are nearly identical at $t=0$ in the middle panel. The total disc mass is shown in the top panel. Aside from the obvious fact that the disc that was fed longer has a much larger mass, the plot clarifies that the dissipation rate of the more massive disc is much slower than the less massive one. This is more easily seen in the middle panel that displays the temporal evolution of the surface density for the two models. Finally, the third panel exhibits the corresponding light curves, demonstrating that the model that had the longer build-up time (and, hence, is more massive) dissipates at a much slower rate than the other model.

In the remainder of the text, we refer to the effect described above as the \textit{mass reservoir effect}, following \cite{rimulo2018}. This realization is important because it points to a fundamental difference between the light curves during disc build-up and dissipation: for the former, the rate at which the brightness varies depends only on $\alpha$ and 
$\dot{J}_{*,\mathrm{std}}$
 while for the latter the rate also depends on how long the disc was fed. C12 did not include the mass reservoir effect in their analysis, as no modeling was done of the build-up phase that preceded the 2004 disc dissipation. As a result, we show below that the $\alpha$ determination in that study was overestimated by a factor of roughly five.


\section {Cycle Lengths, Growth and Decay Rates}
\label{cycle_lengths}

Before discussing the results of the VDD modeling of the light curve of $\omega$ CMa, we present an analysis of the light curve made with the following formula

\begin{equation}
\label{magnitude}
m_\mathrm{v}(t) = (m_{0}-m^{\prime})e^{-\frac{(t-t_{0})}{\tau}}+m^{\prime},
\end{equation}
where $m_\mathrm{v}(t)$ is the visual magnitude as a function of time, $m_{0}$ is the visual magnitude at the beginning of each section of the lighcurve ($t=t_0$), $m^{\prime}$ is the asymptotic visual magnitude of each section, and $\tau$ is the associated time-scale of the brightness variation. The motivation for such an analysis is twofold. First, it allows for obtaining important information about the light curve, such as what are the relevant sections to model and their lengths. Second, as we expect that the rate of brightness variation is controlled mostly by $\alpha$ (see Eq.~\ref{sigmadot}), this analysis will give us hints as to possible temporal variations of this parameter.


\begin{figure}
\centering
\includegraphics[width=1.0\linewidth]{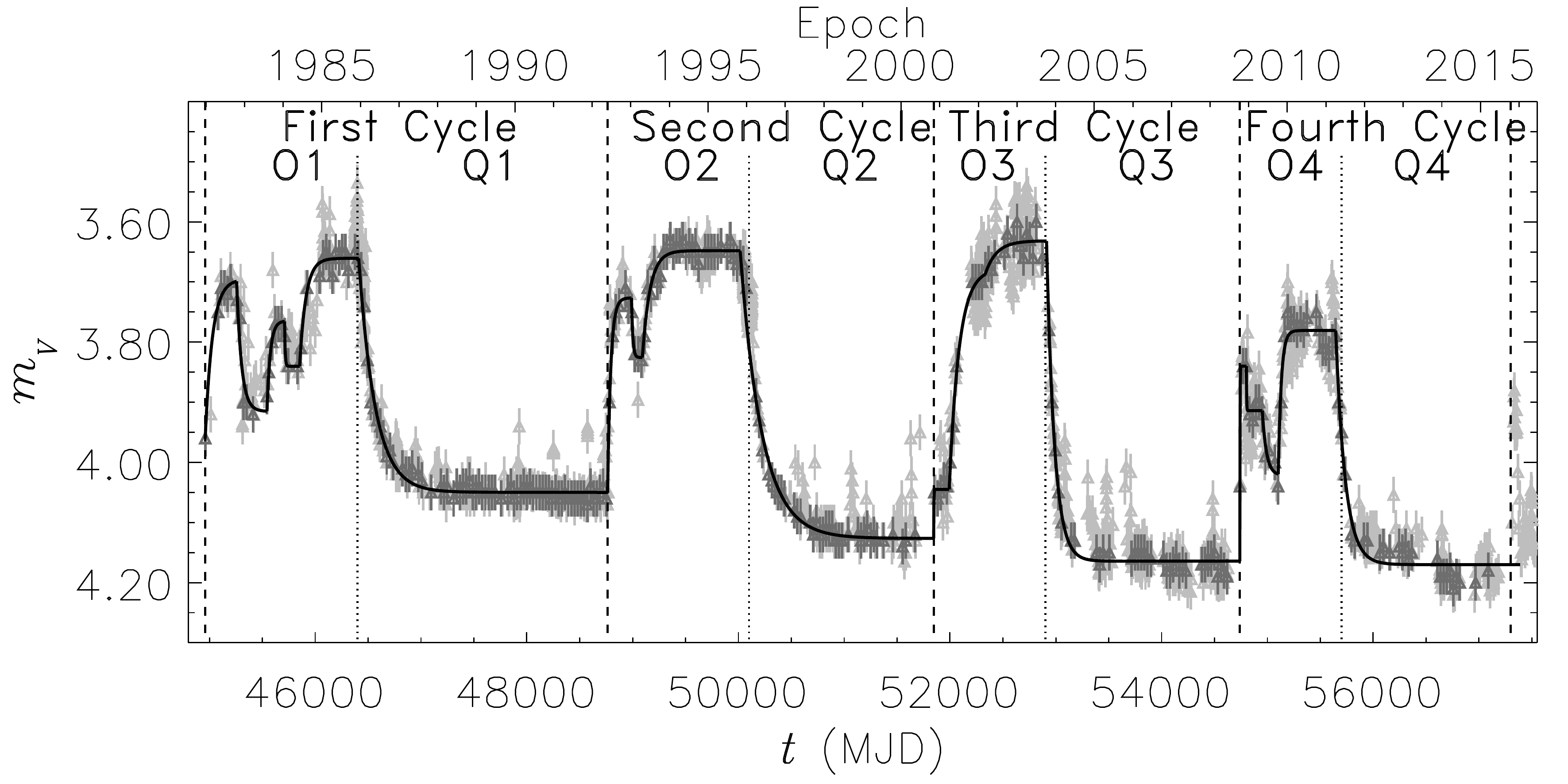}\par\medskip
\caption{The exponential fit of the $V$-band photometry data of $\omega$ CMa (solid line). The light grey triangles show the whole dataset while the dark grey ones show the averaged data in 30-day intervals.}
\label{exp_fit}
\end{figure}



\begin{table*}
\begin{center}
\caption{The parameters derived using exponential formula fitting (columns 2 to 6) and the parameters obtained by the VDD model (columns 7 to 9).}
\begin{tabular}{@{}ccccccccccccccccc}
\hline
\hline
phase & \vline & length & $\vert\Delta m\vert$ & $m_{0}$ & $\tau$ & $t_0^{\rm d}$ & $\alpha$ & $\dot{M}_\mathrm{inj}$ & $-\dot{J}_{*,\mathrm{std}}$ \\

& \vline & (days) & & & (days) & (MJD) & & $(10^{-7}\mathrm{M_\odot yr^{-1}})$ & $(10^{36}\mathrm{g\,cm^2\,s^{-2}})$ \\

\hline  & \vline \\

Cycle 1  & \vline & 3803 & 0.09 & 3.96 & ----- & 44960.5 & ----- & ----- & ----- \\

O1  & \vline & 1451 & 0.30 & 3.96 & ----- & 44960.5 & ----- & ----- & ----- \\

O1$_\mathrm{o1}$  & \vline & 297 & 0.26 & 3.96 & 68.0 & 44960.5 & 1.00 & 3.0 & 4.4 \\

O1$_\mathrm{q1}$  & \vline & 287 & 0.22 & 3.70 & 46.1 & 45257.5 & 1.00 & 1.4 & 2.0 \\

O1$_\mathrm{o2}$  & \vline & 166 & 0.16 & 3.92 & 31.1 & 45544.5 & 1.00 & 2.5 & 3.6 \\

O1$_\mathrm{q2}$  & \vline & 142 & 0.08 & 3.76 & 17.4 & 45710.5 & 1.00 & 1.9 & 2.7 \\

O1$_\mathrm{o3}$  & \vline & 559 & 0.18 & 3.84 & 59.5 & 45852.5 & 1.00 & 3.4 & 4.9 \\

Q1  & \vline & 2352 & 0.39 & 3.66 & 156.3 & 46411.5 & 0.20 $\pm 0.03$ & 0.2 & 0.2 \\

& \vline \\

Cycle 2  & \vline & 3082 & 0.07 & 4.05 & ----- & 48763.5 & ----- & ----- & ----- \\

O2  & \vline & 1253 & 0.40 & 4.05 & ----- & 48763.5 & ----- & ----- & ----- \\

O2$_\mathrm{o1}$  & \vline & 226 & 0.32 & 4.05 & 30.1 & 48763.5 & 1.00 & 3.0 & 4.4 \\

O2$_\mathrm{q1}$  & \vline & 103 & 0.10 & 3.73 & 20.0 & 48989.5 & 1.00 & 1.9 & 2.8 \\

O2$_\mathrm{o2}$  & \vline & 924 & 0.18 & 3.83 & 71.9 & 49092.5 & 1.00 & 3.4 & 4.9 \\

Q2  & \vline & 1829 & 0.48 & 3.65 & 204.1 & 50016.5 & 0.13 $\pm 0.01$ & 4.0$\times$10$^{-3}$ & 5.8$\times$10$^{-3}$ \\

& \vline \\

Cycle 3  & \vline & 2894 & 0.04 & 4.12 & ----- & 51845.5 & ----- & ----- & ----- \\

O3  & \vline & 1067 & 0.49 & 4.12 & ----- & 51845.5 & ----- & ----- & ----- \\

O3$_\mathrm{o1}$  & \vline & 147 & 0.08 & 4.12 & 16.3 & 51845.5 & 1.00 & 3.7 & 5.4 \\

O3$_\mathrm{o2}$  & \vline & 339 & 0.35 & 4.04 & 92.6 & 51992.5 & 0.10 & 0.4 & 0.6 \\

O3$_\mathrm{o3}$  & \vline & 581 & 0.06 & 3.69 & 107.5 & 52331.5 & 0.10 & 0.4 & 0.6 \\

Q3  & \vline & 1827 & 0.53 & 3.63 & 74.6 & 52912.5 & 0.21 $\pm 0.05$ & 4.0$\times$10$^{-3}$ & 5.8$\times$10$^{-3}$ \\

& \vline \\

Cycle 4  & \vline & 2561 & 0.03 & 4.16 & ----- & 54739.5 & ----- & ----- & ----- \\

O4  & \vline & 906 & 0.38 & 4.16 & ----- & 54739.5 & ----- & ----- & ----- \\

O4$_\mathrm{o1}$  & \vline & 62 & 0.32 & 4.16 & 15.2 & 54739.5 & 1.00 & 2.6 & 3.7 \\

O4$_\mathrm{q1}$  & \vline & 150 & 0.07 & 3.84 & 15.8 & 54801.5 & 1.00 & 1.6 & 2.3 \\

O4$_\mathrm{q2}$  & \vline & 151 & 0.11 & 3.91 & 39.4 & 54951.5 & 0.10 & 0.1 & 0.1 \\

O4$_\mathrm{o2}$  & \vline & 388 & 0.26 & 4.02 & 33.8 & 55102.5 & 1.00 & 2.6 & 3.8 \\

O4$_\mathrm{q3}$  & \vline & 155 & 0.05 & 3.76 & 17.1 & 55490.5 & 0.50 & 1.0 & 1.4 \\

Q4  & \vline & 1655 & 0.39 & 3.81 & 85.5 & 55645.5 & 0.11 $\pm 0.03$ & 2.0$\times$10$^{-3}$ & 2.9$\times$10$^{-3}$ \\

\hline uncertainty & \vline & $\pm 1$ & $\pm 0.02$ & $\pm 0.01$ & $\pm 0.5$ & $\pm 0.5$ & ----- & ----- & ----- \\

\hline
\end{tabular}
\label{results}
\end{center}
\end{table*}


As seen in Sect.~\ref{observations}, the light curve of $\omega$ CMa is composed of major outburst and quiescence phases over which several small-scale phases (flickers) are seen. The major phases are long-lived (a few months and longer) while the flickers happen over much shorter timescales. Since modeling all the phases in the light curve would be a very demanding task, we concentrate on fitting only the major phases, which are (arbitrarily) defined as the phases either longer than 3 months, and with magnitude variations larger than 0.05 or longer than 2 months, and with magnitude variations larger than 0.30. Selecting the desired phases was done after averaging the data in 30-day intervals as the resulted data formed a smoother light curve (the dark grey triangles in Fig.~\ref{exp_fit}). This choice of focusing on the major phases is justified by i) considering that the short-term phases contribute little to the total mass budget of the disc, and ii) being short-lived, the short-term phases are usually poorly sampled by the observations. The important point is that by modeling only the major phases we have a good handle on the total disc mass at the end of build-up, which is important to model the dissipation phase because the disc dissipation is history dependent (Sect.~\ref{mass_reservoir}). 

The results of the exponential fitting are shown in Fig.~\ref{exp_fit}, and the list of sections considered with their respective parameters are presented in Table~\ref{results}. The derived cycle lengths are also indicated in Fig.~\ref{overlapped} by the horizontal arrows. The lengths of successive formation phases and successive dissipation phases were found to be decreasing, as anticipated in Sect.~\ref{observations} by eye inspection. This phenomenon remains unexplained. On the other hand, it can be seen that there is a negative correlation between the length of formation and the rate of disc dissipation, as expected from the mass reservoir effect.

Finally, we found that the growth and decay time-scales of the light curve, $\tau$, vary from cycle to cycle and also within a given cycle (Table~\ref{results}). This result suggests that the viscosity parameter, $\alpha$, is varying, as $\alpha$ is the main parameter controlling the disc evolution time-scales. The selected phases, as well as their starting times and duration, will be used in the next section as input for the VDD modeling of the light curve of $\omega$ CMa.


\section {Results of the application of Viscous Decretion Disc model}
\label{vdd_model}

One of the most striking features of $\omega$ CMa's light curve (Fig.~\ref{complete_lightcurve}) is that the star always presents a flux excess, if one assumes that the central star is intrinsically non-variable above the $\sim 0\fm05$ level. This is a reasonable assumption, because known mechanisms that could cause the star to vary in brightness account only for variations of much lower amplitudes. For instance, pulsations could account only for short-period (days or less) variations at the tens of mmag level (e.g. \citealt{baade2016}; \citealt{kurtz2015}; \citealt{balona2011}; \citealt{huat2009}). In the following we therefore assume that a variable $V$-band excess was present in the past 34 years, and this excess is of disc origin.

\begin{figure}
\centering
\includegraphics[width=1.0\linewidth]{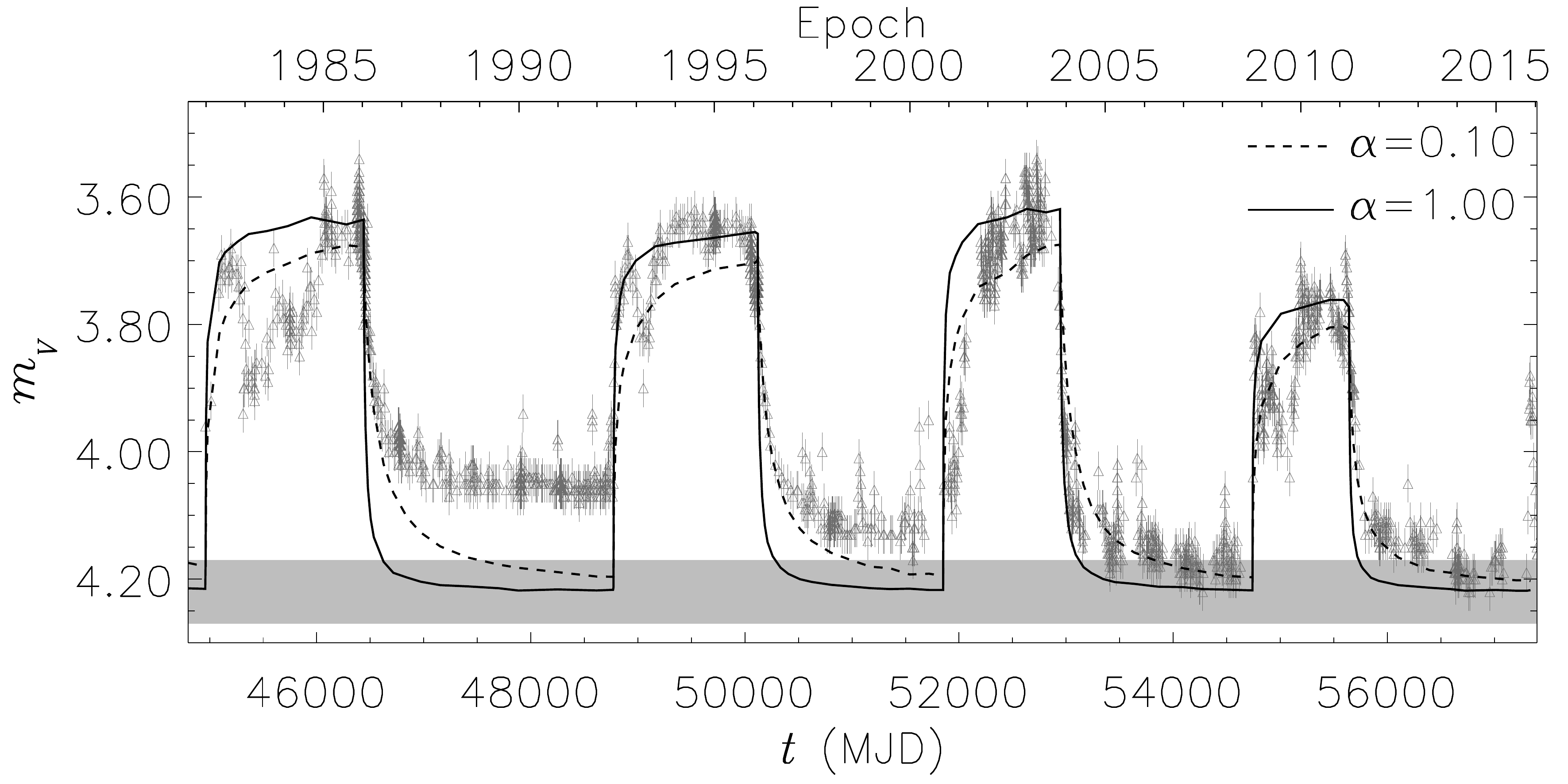}
\caption{Initial VDD fitting of $\omega$ CMa's light curve. At outburst, different mass injections were assumed in order to fit the brightness level at the end of outburst. The phases of disc dissipation were assumed to have $-\dot{J}_{*,\mathrm{std}}=0$. Each line type corresponds to different values of viscosity parameter, as indicated. The data are the same as Fig.~\ref{complete_lightcurve} (also for the Figs.~\ref{vdd_model_c1} to \ref{summary}).}
\label{vdd_initial}
\end{figure}

The prevalence of a flux excess poses a modeling difficulty illustrated in Fig.~\ref{vdd_initial}. In this figure, only a coarse fitting of the past four cycles was attempted. Each cycle was assumed to consist of a single outburst, during which the disc mass injection had a non-zero value, and a single quiescence, for which $-\dot{J}_{*,\mathrm{std}}=0$. Furthermore, the viscosity was assumed to be constant throughout the four cycles. In Fig.~\ref{vdd_initial}, models for two values of $\alpha$ are displayed: 0.1 and 1.0. The values of $-\dot{J}_{*,\mathrm{std}}$ adopted in the models are listed in Table~\ref{jdot_steady_value}. The calculations were performed using the hydrodynamic models described in Sect.~\ref{model}.


\begin{table}
\begin{center}
\caption{Adopted values of $-\dot{J}_{*,\mathrm{std}}$ for the models shown in Fig.~\ref{vdd_initial}.}
\begin{tabular}{@{}ccc}
\hline
\hline
& \multicolumn{2}{c}{$-\dot{J}_{*,\mathrm{std}}\,(10^{36}\mathrm{g\,cm^2\,s^{-2}})$} \\ 
 & $\alpha = 0.1$ & $\alpha = 1.0$ \\
\hline
$\mathrm{O1}$ & 0.357 & 3.57 \\
$\mathrm{O2}$ & 0.339 & 3.39 \\
$\mathrm{O3}$ & 0.386 & 3.86 \\
$\mathrm{O4}$ & 0.250 & 2.50 \\
\hline
\end{tabular}
\label{jdot_steady_value}
\end{center}
\end{table}


The models in Fig.~\ref{vdd_initial} are clearly unable to reproduce the data. The single outburst assumption fails to reproduce the complex observed light curve, as expected. In the high-viscosity models the inner disc quickly dissipates, and the flux excess quickly goes to zero, in stark contrast to the observations. Low-viscosity models perform no better: they fail to reproduce the initial phase of disc dissipation (having a too slow flux variation), and also the phase of nearly-constant flux that is reached a few months after dissipation started. This simple model also cannot offer any explanation for the secular fading observed at the end of each quiescence phase.

In order to solve these issues, we propose an alternate scenario for the dimming phases of $\omega$ CMa, in which $-\dot{J}_{*,\mathrm{std}}$ may be different than zero, as is usually assumed. Therefore, $-\dot{J}_{*,\mathrm{std}}$ can have any physically sound value with larger $-\dot{J}_{*,\mathrm{std}}$ for outburst phases during which the disc attains a higher density, and smaller $-\dot{J}_{*,\mathrm{std}}$ for dissipation phases. In the transition between outburst and dissipation, the disc would therefore switch between a high density phase to a low density one. In other words, the dimmings observed in $\omega$ CMa would be partial, rather than full, disc dissipations. 

In the following, we present a refinement of the model presented in Fig.~\ref{vdd_initial}. The modeling assumptions are:
\begin{itemize} 
\item The sections defined in Table~\ref{results} are modeled individually and in temporal succession. 
\item For each section of the light curve, the free parameters are the viscosity parameter $\alpha$, the disc mass injection rate, parameterized by $-\dot{J}_{*,\mathrm{std}}$, and the starting time of the section.
\item During dissipation, $-\dot{J}_{*,\mathrm{std}}$ may have a non-zero value.
\item Since the observations prior to 1982 are scarce, we did not attempt to model them. However, as observations clearly indicate the presence of a disc prior to 1982 (see Fig.~\ref{complete_lightcurve}), the model starts with a steady-state disc followed by a 5-year long dissipation that ends when O1 begins. The choice of 5-year dissipation is justified as this is the typical length of the observed dissipations. The assumed AM injection rate for this previous phase is consistent with the value assumed for the first cycle.
\item \cite{maintz2003} did not provide estimates for the uncertainties in the stellar parameters and although variations in the stellar parameters might affect our results, we trust their values for modeling simplification.
\end{itemize} 


\subsection {Results for the first cycle}
\label{vdd_c1}

Modeling started by considering a single value of $\alpha = 1.0$ for the first part of O1. 
Other possible values of $\alpha$ were not examined because the data for this part are very sparse. This small-scale outburst (O1$_\mathrm{o1}$) was modeled as the continuation of a 5-year long quiescence that started with a steady-state disc (see above).

\begin{figure*}
\begin{minipage}{0.495\linewidth}
\centering
{\includegraphics[width=1.0\linewidth, height=0.5\linewidth]{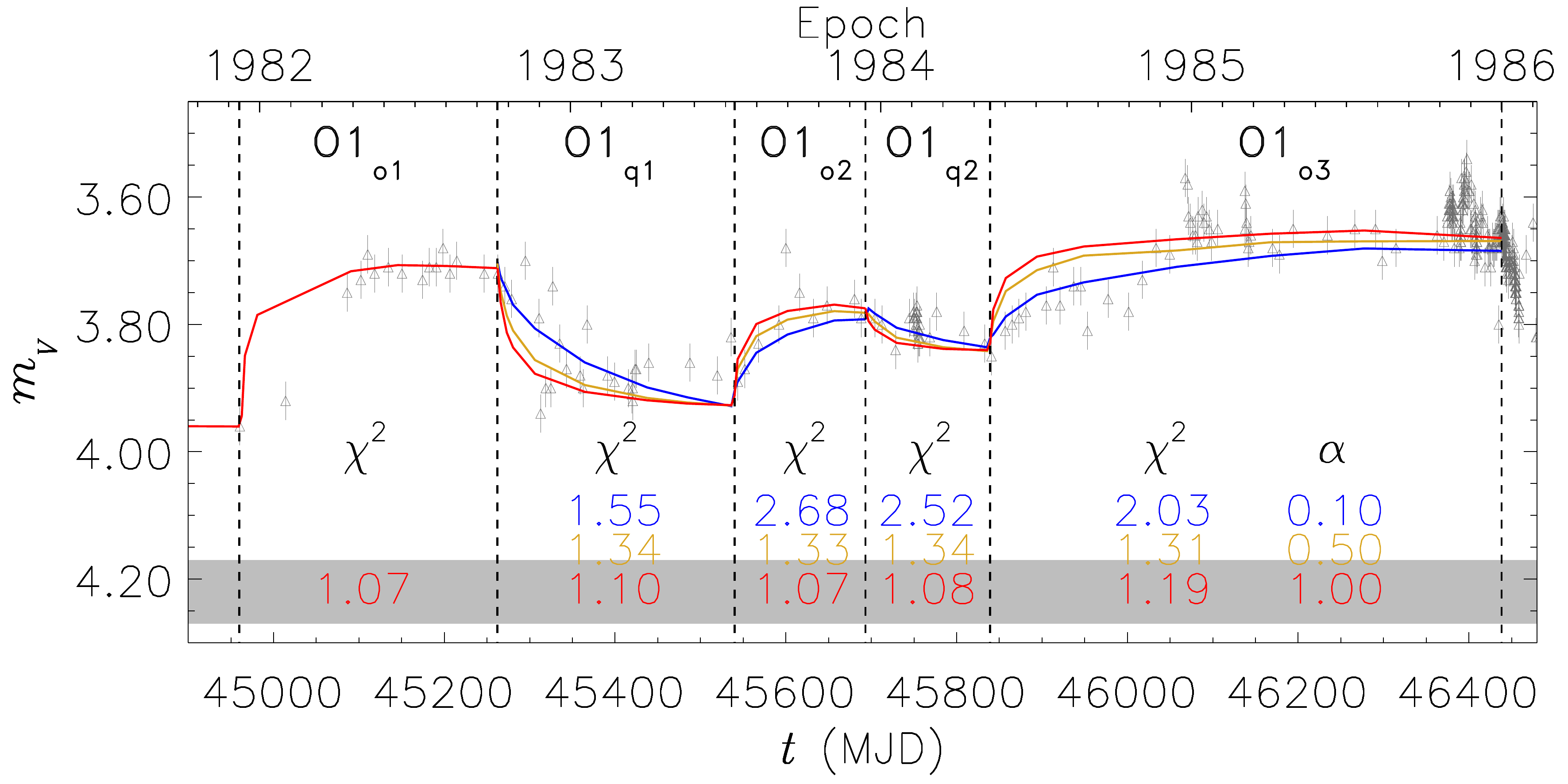}}
\end{minipage}
\begin{minipage}{0.495\linewidth}
\centering
{\includegraphics[width=1.0\linewidth, height=0.5\linewidth]{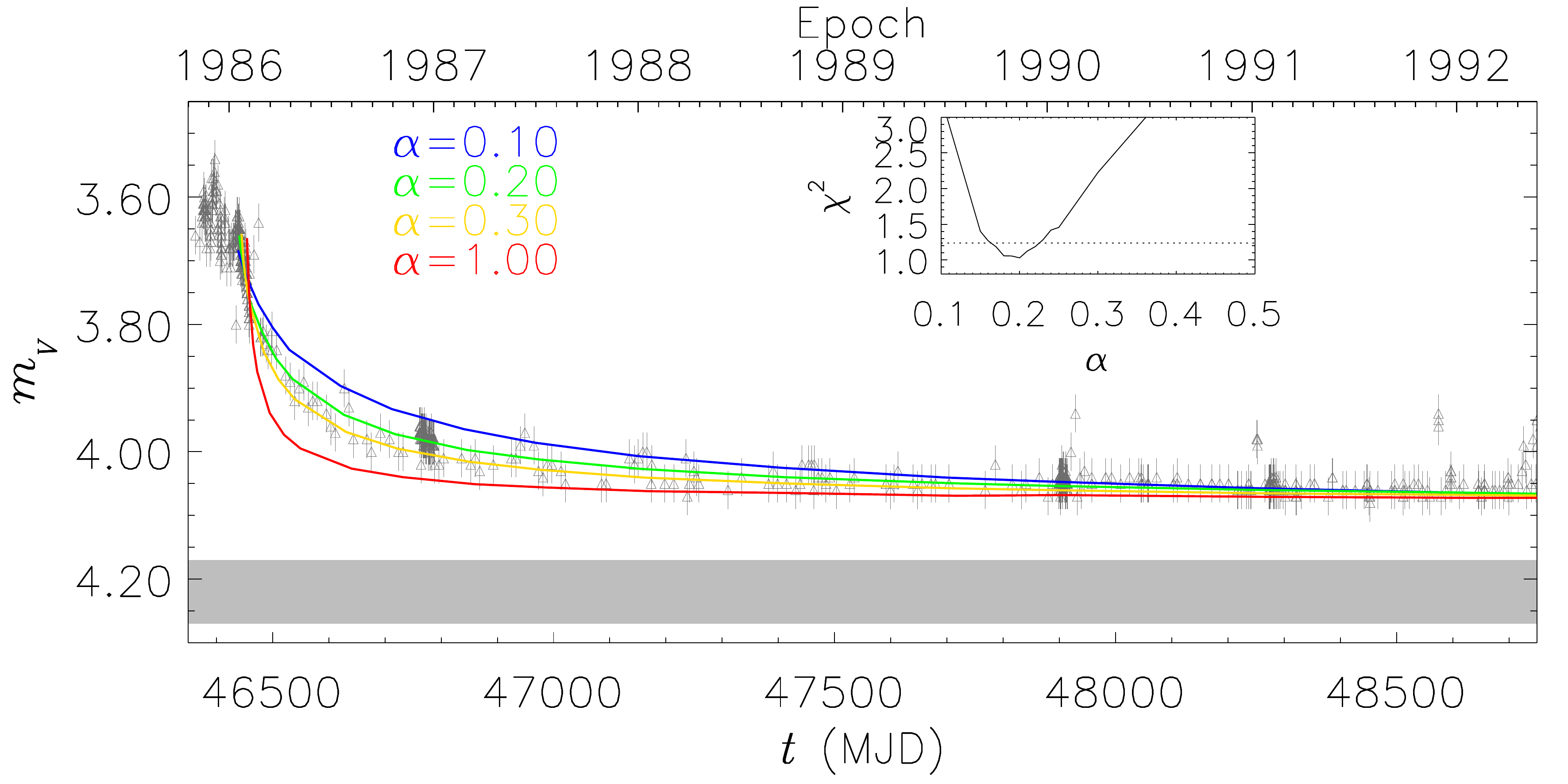}}
\end{minipage}
\caption{
$V$-band light curve of $\omega$ CMa for the first cycle (points). 
{\it Left}: Models for O1 for three different values of $\alpha$: 0.10, 0.50 and 1.00, as indicated. 
The $\chi^2_\mathrm{red}$ for each model and each section the light curve is indicated.
{\it Right}: Model for Q1 for four values of $\alpha$: 0.10, 0.20, 0.30, and 1.00, as indicated. The estimated best-fit value of $\alpha = 0.20 \pm 0.03$. The $\chi^2_\mathrm{red}$ for different values of $\alpha$ is shown in the inset and the 90\% confidence level is indicated with the horizontal dotted line. The horizontal grey band represents the estimated intrinsic visual magnitude of the central star of $\omega$ CMa.
}
\label{vdd_model_c1}
\end{figure*}

For the remaining sections of O1, which alternate build-up with dissipation phases, we explored three different values of $\alpha$ (0.10, 0.50, and 1.00 -- Fig.~\ref{vdd_model_c1}, left panel). For the dissipation phases, the criterium used was to favor models that matched better the lower points in the light curves, as the higher points likely result from small-scale, poorly-sampled outbursts (flickers) on top  of the dissipation. The reduced chi-squared values, $\chi^2_\mathrm{red}$, for each model, shown in the plot, clearly indicate that the $\alpha=1.0$ model better fits all sections of O1. A complex AM injection rate is necessary to reproduce the general behavior of the light curve in O1. A graphical representation of the AM injection rate is shown in the third panel of Fig.~\ref{summary}.

The dissipation phase that followed O1 (Q1) has a much smoother shape, with only a handful of small-scale flickers, which greatly facilitates the modeling. It is possible that the smoother shape during dissipation is simply a result of the lower mass injection rate: when the star is less noisy, its mean mass injection rate is also lower, and, as a result, the disc dissipates. We determined that $-\dot{J}_{*,\mathrm{std}}$ decreases from $4.9 \times 10^{36}\mathrm{g\,cm^2\,s^{-2}}$ at the end of O1 to $2.1 \times 10^{35}\mathrm{g\,cm^2\,s^{-2}}$ at the end of Q1, causing an overall density decrement (see Fig.~\ref{summary}) that explains the observed decrease in the disc excess ($\Delta V = 0.39$). We found a best-fit value of $\alpha = 0.20 \pm 0.03$, shown in the right panel of Fig.~\ref{vdd_model_c1}, along with models for other values of $\alpha$ for comparison. The inset shows the $\chi^2_\mathrm{red}$ of the fit as a function of $\alpha$. The best fit value was determined from the minimization of the $\chi^2_\mathrm{red}$ value, for which we adopted the following definition:
\begin{equation}
\label{chisquare}
\chi^2_\mathrm{red} = \displaystyle\sum_{i=1}^{N} \frac{(m\mathrm{^{obs}_{v,\it{i}}} -m\mathrm{^{mod}_{v,\it{i}}})^2}{\sigma^2_{i}(N-2)},
\end{equation}
where $(m\mathrm{^{obs}_{v,\it{i}}}$ and $m\mathrm{^{mod}_{v,\it{i}}}$ are the observed and model magnitude, respectively, $\sigma_{i}$ is the observed magnitude error, and $N$ is the total number of data points that were fitted. The best $\alpha$ value is found at the minimum $\chi^2_\mathrm{red}$ function, which was computed for a grid with a step size of 0.01. The uncertainty of this value was estimated from the $\Delta\chi^2_\mathrm{red, 0.90} = \chi^2_\mathrm{red} - \chi^2_\mathrm{red, min}$ intersections, where $\Delta\chi^2_\mathrm{red, 0.90}$ is a function of the number of degrees of freedom of the fit for 90\% of confidence \citep[see Chapter~11 of][]{bevington1992}. For all the cases, the derived uncertainty value was found to be larger than the adopted step size, and therefore the $\chi^2_\mathrm{red}$ function was sufficiently well sampled.


\subsection {Results for the second cycle}
\label{vdd_c2}

The model fitting of the second cycle followed the same procedure as presented for the previous cycle. 
In many ways, the second cycle is similar to the first one. As before, O2 is composed of alternating phases of build-up and dissipation, which implies a complex disc feeding history. A short dissipation phase (O2$_\mathrm{q2}$) precedes the long dissipation phase Q2.

\begin{figure*}
\begin{minipage}{0.495\linewidth}
\centering
{\includegraphics[width=1.0\linewidth]{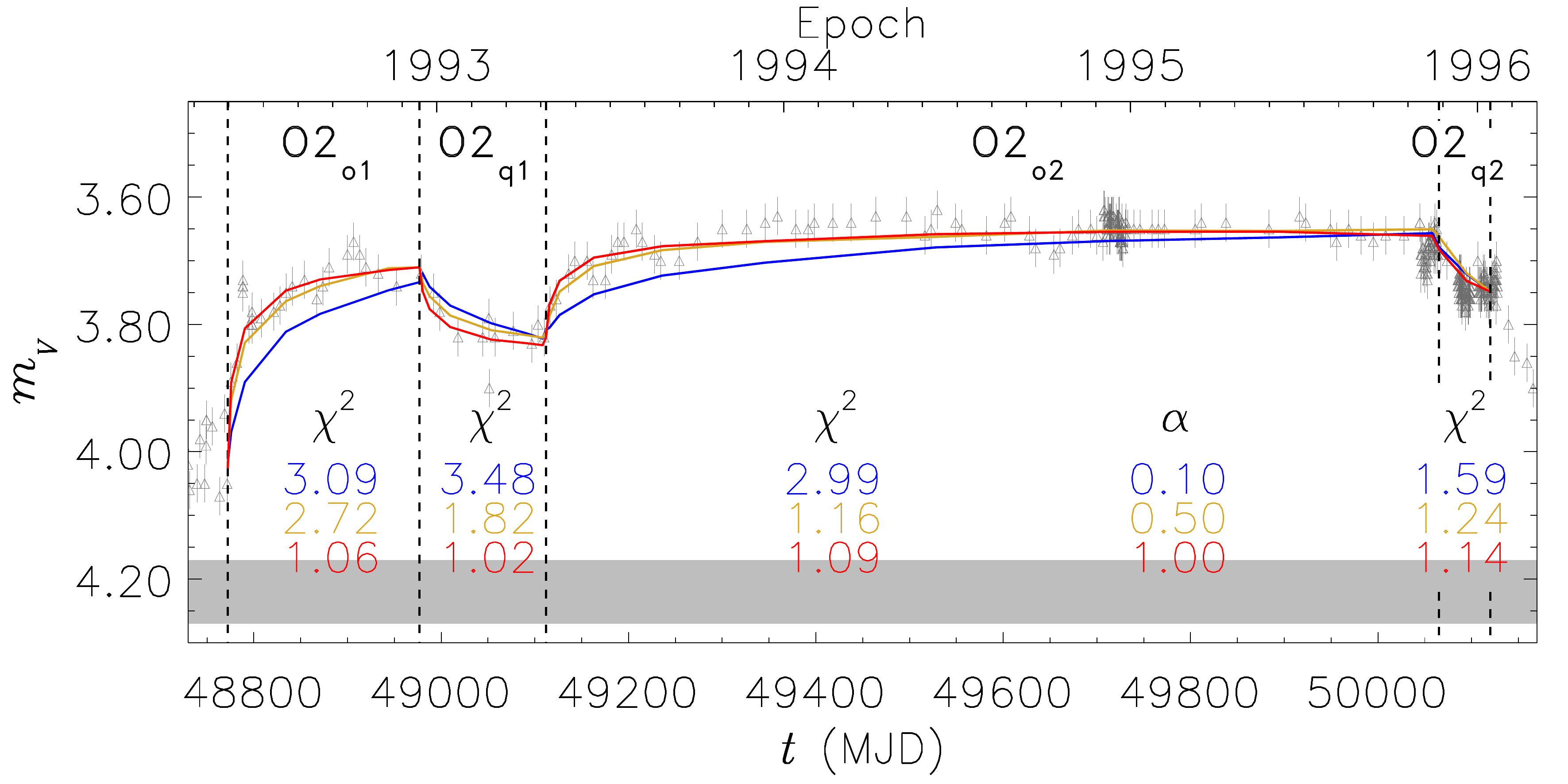}}
\end{minipage}
\begin{minipage}{0.495\linewidth}
\centering
{\includegraphics[width=1.0\linewidth]{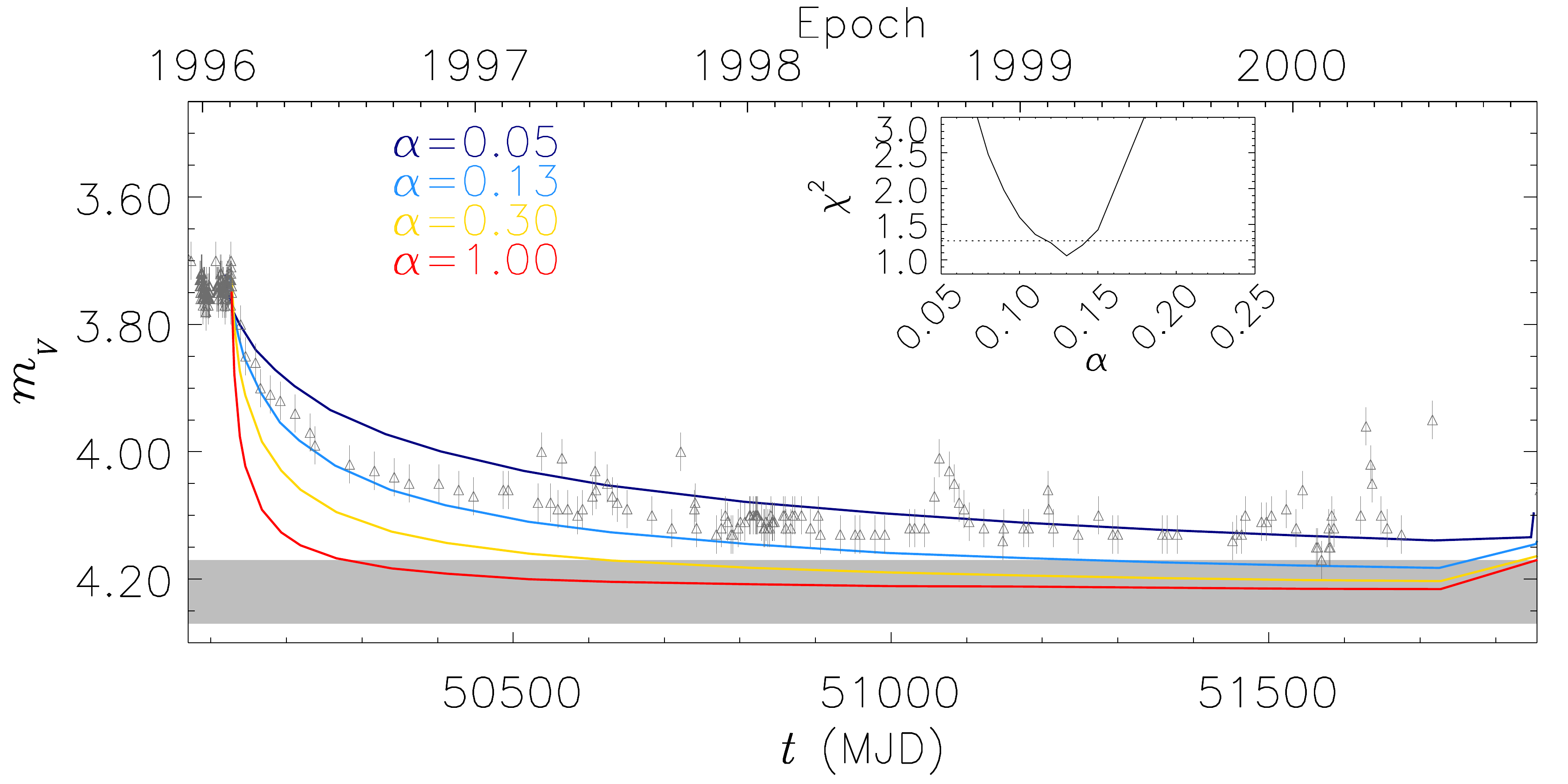}}
\end{minipage}
\caption{Same as Fig.~\ref{vdd_model_c1} for the second cycle. The best-fit value for the outburst is $\alpha = 1.0$ and for the dissipation, $\alpha = 0.13 \pm 0.01$.}
\label{vdd_model_c2}
\end{figure*}

Again, a value of $\alpha = 1.0$ seems to best represent the entire O2 phase. For Q2, we estimate $\alpha = 0.13 \pm 0.01$ and $-\dot{J}_{*,\mathrm{std}} = 5.8 \times 10^{33}\mathrm{g\,cm^2\,s^{-2}}$. The fit model is shown in Fig.~\ref{vdd_model_c2}.


\subsection {Results for the third cycle}
\label{vdd_c3}

\begin{figure*}
\begin{minipage}{0.495\linewidth}
\centering
{\includegraphics[width=1.0\linewidth]{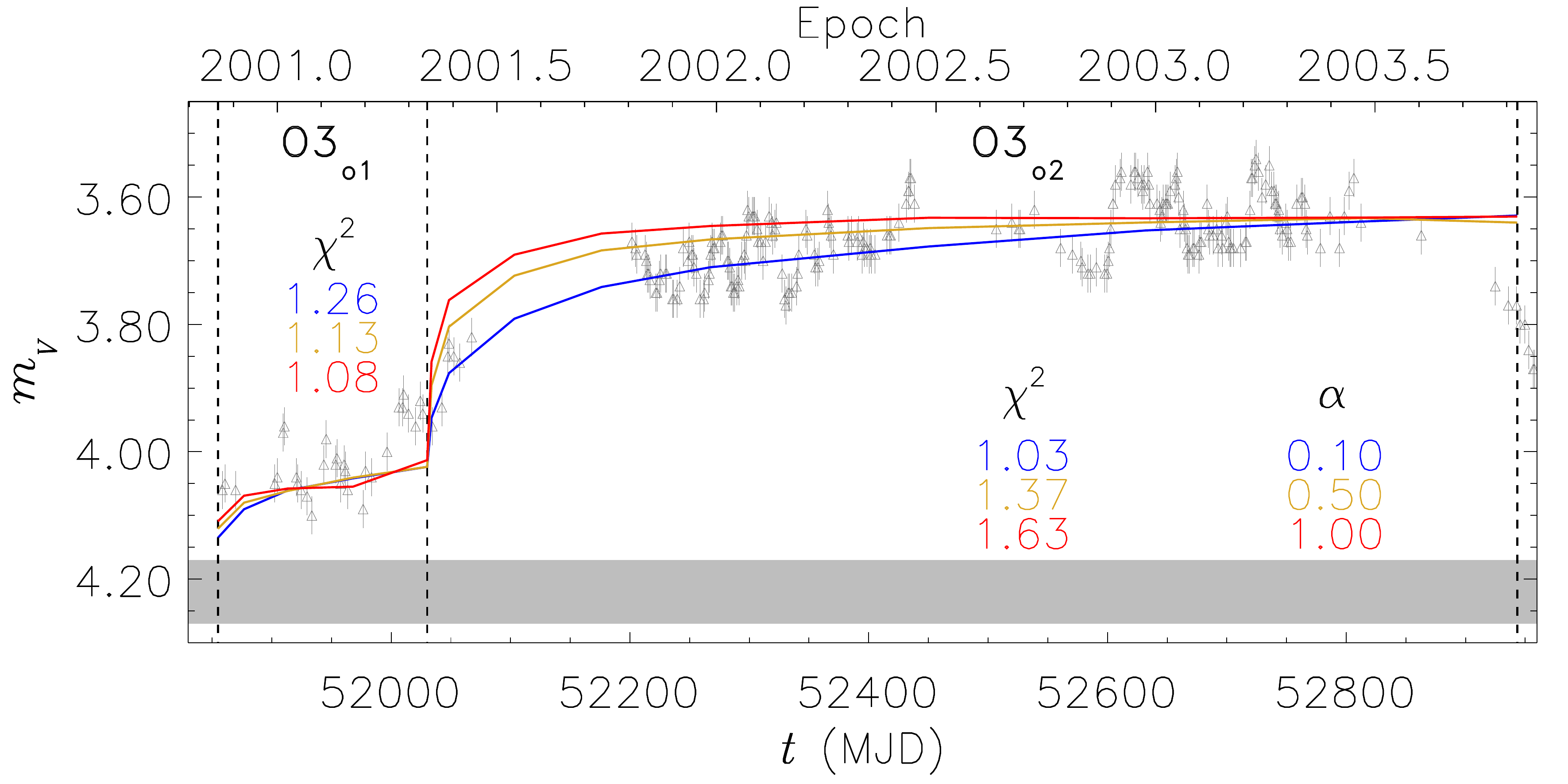}}
\end{minipage}
\begin{minipage}{0.495\linewidth}
\centering
{\includegraphics[width=1.0\linewidth]{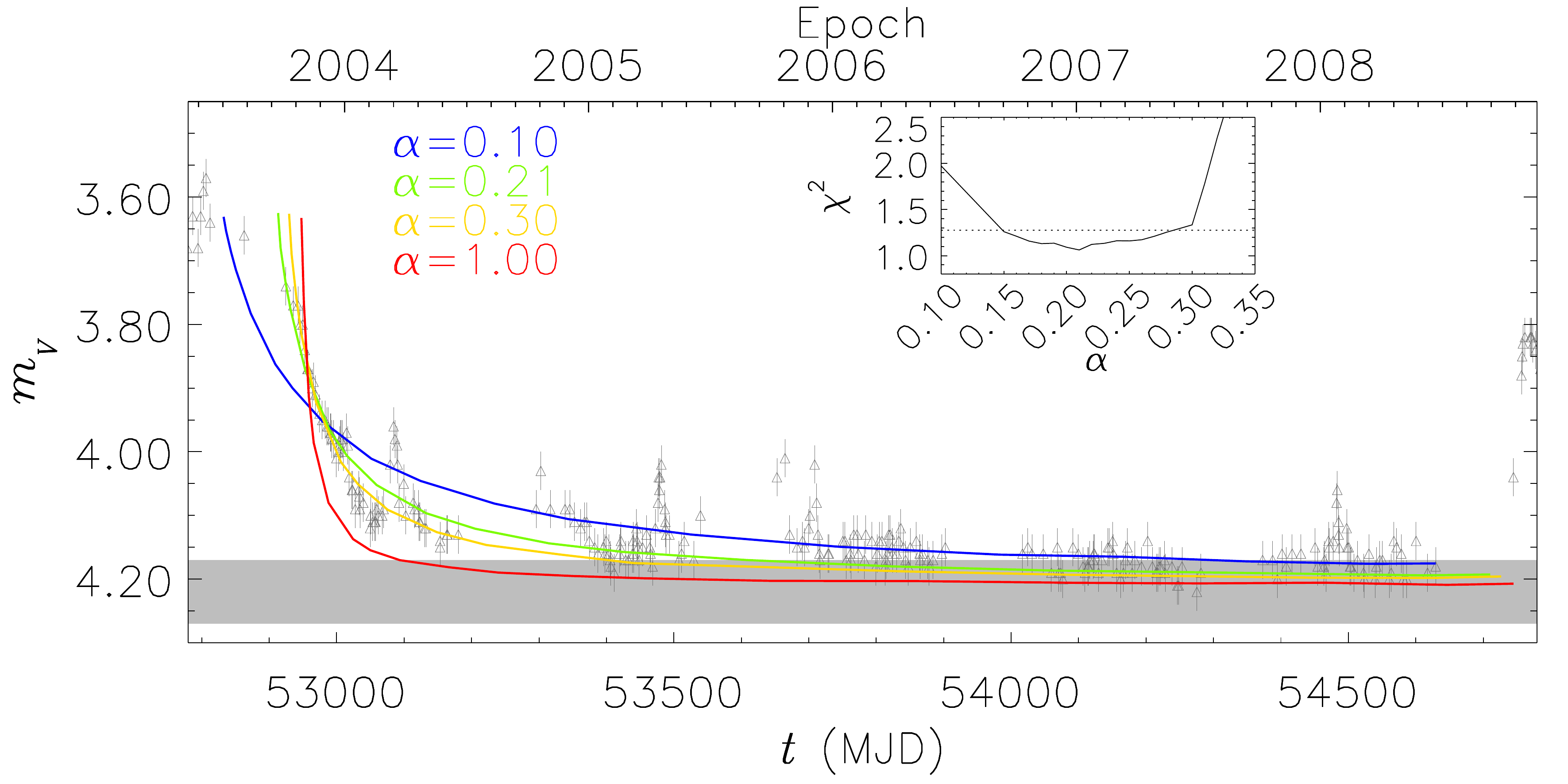}}
\end{minipage}
\caption{Same as Fig.~\ref{vdd_model_c1} for the third cycle. The best fit values were $\alpha = 1.0$ for O3$_\mathrm{o1}$, $\alpha = 0.1$ for O3$_\mathrm{o2}$, and  $\alpha = 0.21 \pm 0.05$ for Q3.
}
\label{vdd_model_c3}
\end{figure*}

For the build-up phases of the third cycle (O3), we again explored $\alpha$ values of 0.10, 0.50 and 1.00 (Fig.~\ref{vdd_model_c3}, left panel). The minimum $\chi^2_\mathrm{red}$ values indicated that $\alpha=1.00$ fits better the O3$_\mathrm{o1}$ section, while $\alpha=0.10$ provides a better fit for the O3$_\mathrm{o2}$ section. However, we believe this last value to be of little statistical significance, as the initial phase of O3$_\mathrm{o2}$, when the largest brightness variations happen, is poorly sampled. Furthermore, O3$_\mathrm{o2}$ displays a quite complex behavior, with many small-scale flickers.
As was the case for O1, a complex AM injection rate is needed to explain the photometric behavior of O3, with $-\dot{J}_{*,\mathrm{std}}$ ranging between $6.1 \times 10^{35}\mathrm{g\,cm^2\,s^{-2}}$ and $5.4 \times 10^{36}\mathrm{g\,cm^2\,s^{-2}}$.

Q3 is of particular relevance because it was previously studied by C12 who found $\alpha = 1.0 \pm 0.2$, which is about 5 times larger than our result of $\alpha=0.21\pm 0.05$ (Fig.~\ref{vdd_model_c3}, right panel). This large discrepancy deserves a careful examination.

\begin{figure}
\centering
\includegraphics[width=1.0\linewidth]{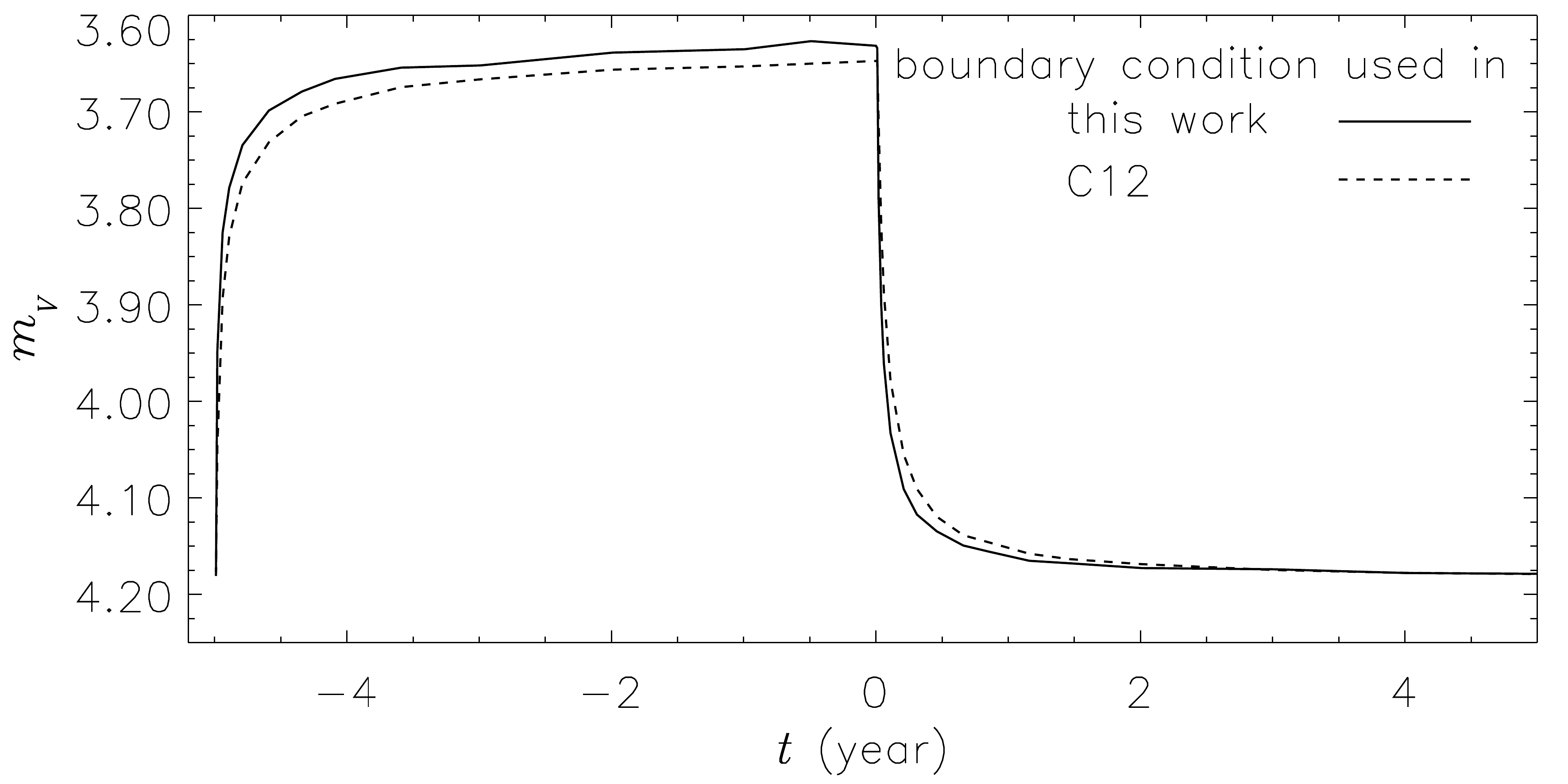}
\caption{Effect of boundary conditions in the {\tt SINGLEBE} code. The dashed line represents the model in which the inner boundary condition is inside the star (used by C12) and the solid line shows the model in which the boundary condition is at the stellar equator that we used. The dashed line exhibits a slower dissipation phase for an equal value of $\alpha$ used in the other model.}
\label{boundary}
\end{figure}

One reason behind the discrepancy between our results for Q3 and C12's is the different boundary conditions used in {\tt SINGLEBE}: C12 assumed $\bar{r}_\mathrm{inj} = 1$ and that the inner boundary condition was inside the star (at $r=0.85\,R_\mathrm{eq}$), while we assumed {$\bar{r}_\mathrm{inj} = 1.02$ and that the inner boundary condition is at the stellar equator}. Figure~\ref{boundary} illustrates the effect of this change in the boundary condition, by comparing two equivalent models with the two different boundary conditions. The boundary conditions used in this work result in a faster disc dissipation, which in turn means that the  $\alpha$ required to match the observed rate of dissipation is smaller. However, this is a small effect, and can explain differences of only $\sim$30$\%$ between the results.
 
The main reason for the discrepancy between the results of C12 and ours is the mass reservoir effect (Sect.~\ref{mass_reservoir}). C12 modeled Q3 by assuming a long previous build-up phase. Therefore, their model overestimated the disc mass at the onset of Q3, which, in turn, caused $\alpha$ to be overestimated. Our results do not suffer from this issue, as we properly took the previous history into account.


\subsection {Results for the fourth cycle}
\label{vdd_c4}

\begin{figure*}
\begin{minipage}{0.495\linewidth}
\centering
{\includegraphics[width=1.0\linewidth]{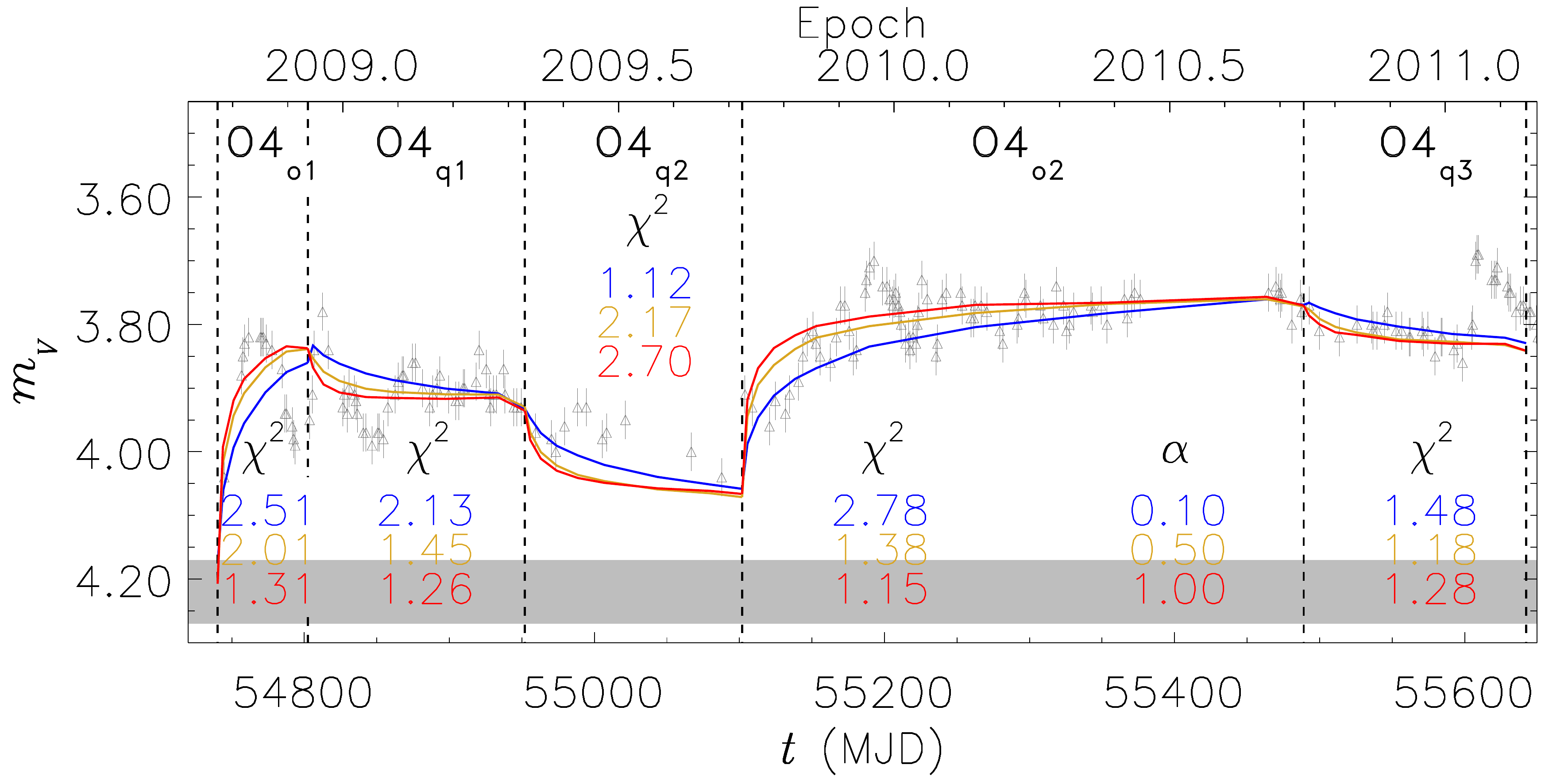}}
\end{minipage}
\begin{minipage}{0.495\linewidth}
\centering
{\includegraphics[width=1.0\linewidth]{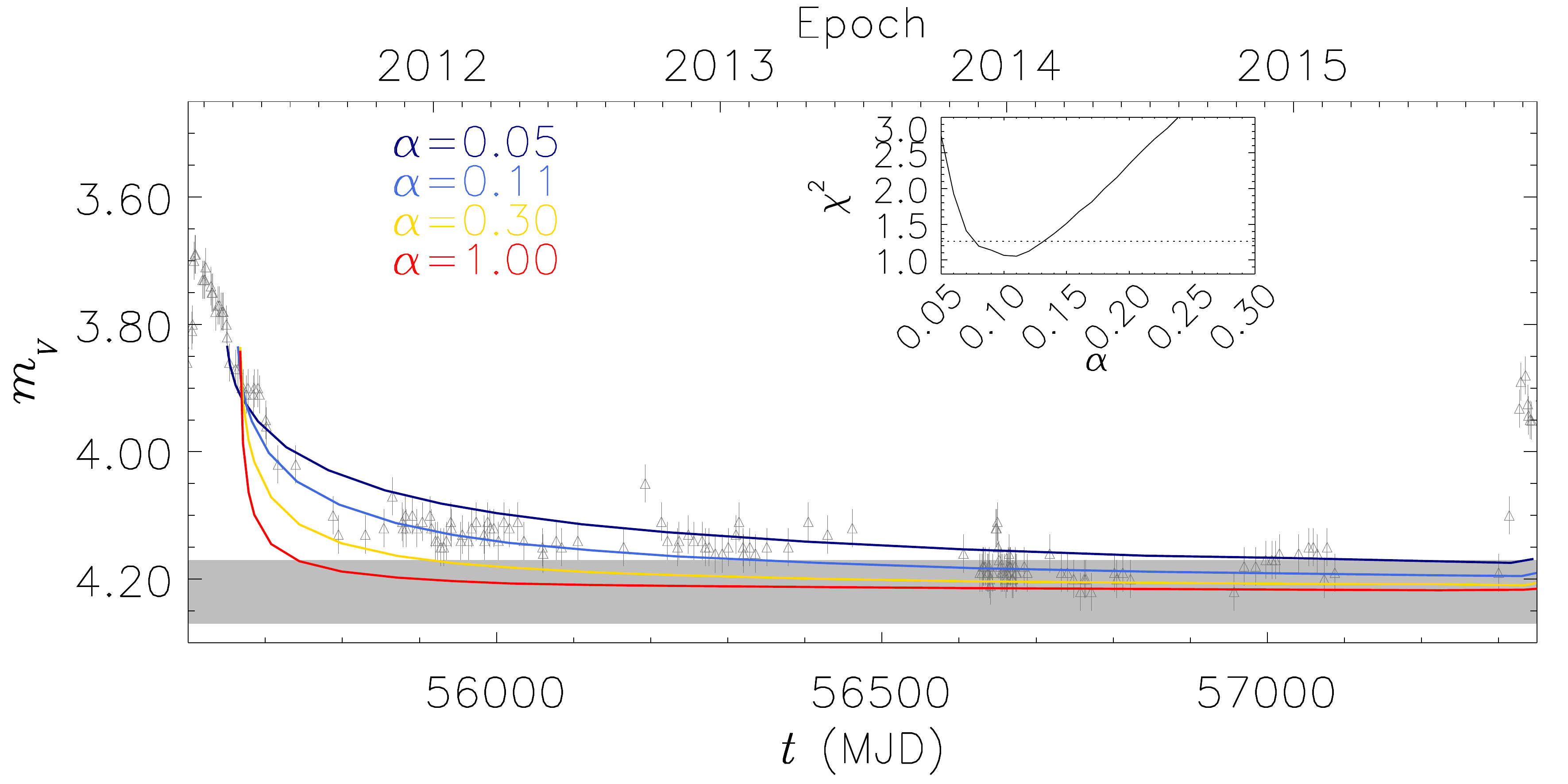}}
\end{minipage}
\caption{Same as Fig.~\ref{vdd_model_c1} for the forth cycle. The best fit for O4 was obtained with $\alpha$ = 1.00 for the first, second, and fourth sections and $\alpha$ = 0.10 and 0.50 for third and fifth sections. For the dissipation phase, $\alpha = 0.11 \pm 0.03$.}
\label{vdd_model_c4}
\end{figure*}

As before, to study O4 we explored three values of $\alpha$ for each section. O4 displays the most complex behavior of all the outbursts, with rapid switches between brightenings and fadings. As a result of our modeling assumptions, according to which we do not attempt to model phases shorter than 60 days, the model fails at reproducing the detailed behavior of the light curve. Figure~\ref{vdd_model_c4}, left panel, shows how each model compares with the data. Formally, the $\chi^2_\mathrm{red}$ values of each section suggest that $\alpha$ may be varying during the outburst from 0.1 to 1.0.

However, we believe these results should be viewed with some caution, as they may depend on the particular choice for the beginning and end of each section. The important point to emphasize is that the modeling, irrespective of the particular choice of $\alpha$ reproduces the general behavior of the light curve, which is enough to estimate the total disc mass at the end of O4. As previously discussed, knowing the total disc mass at the end of the outburst is necessary to properly model the subsequent dissipation phase. The complex behavior of O4 is suggestive of a quite complex AM injection history, as shown in Fig.~\ref{summary}.

The model for Q4 results in $\alpha = 0.11 \pm 0.01$ and  $-\dot{J}_{*,\mathrm{std}} = 2.8 \times 10^{33}\mathrm{g\,cm^2\,s^{-2}}$. It is interesting to note that even though small, this $-\dot{J}_{*,\mathrm{std}}$ is required to reproduce the slight excess still present at the end of Q4 under the hypothesis of reliable stellar parameters.


\section {Discussion}
\label{discussion}


\subsection{The life cycles of $\omega$ CMa's disc}
\label{life_cycle}

\begin{figure*}
\centering
{\includegraphics[width=1.0\linewidth]{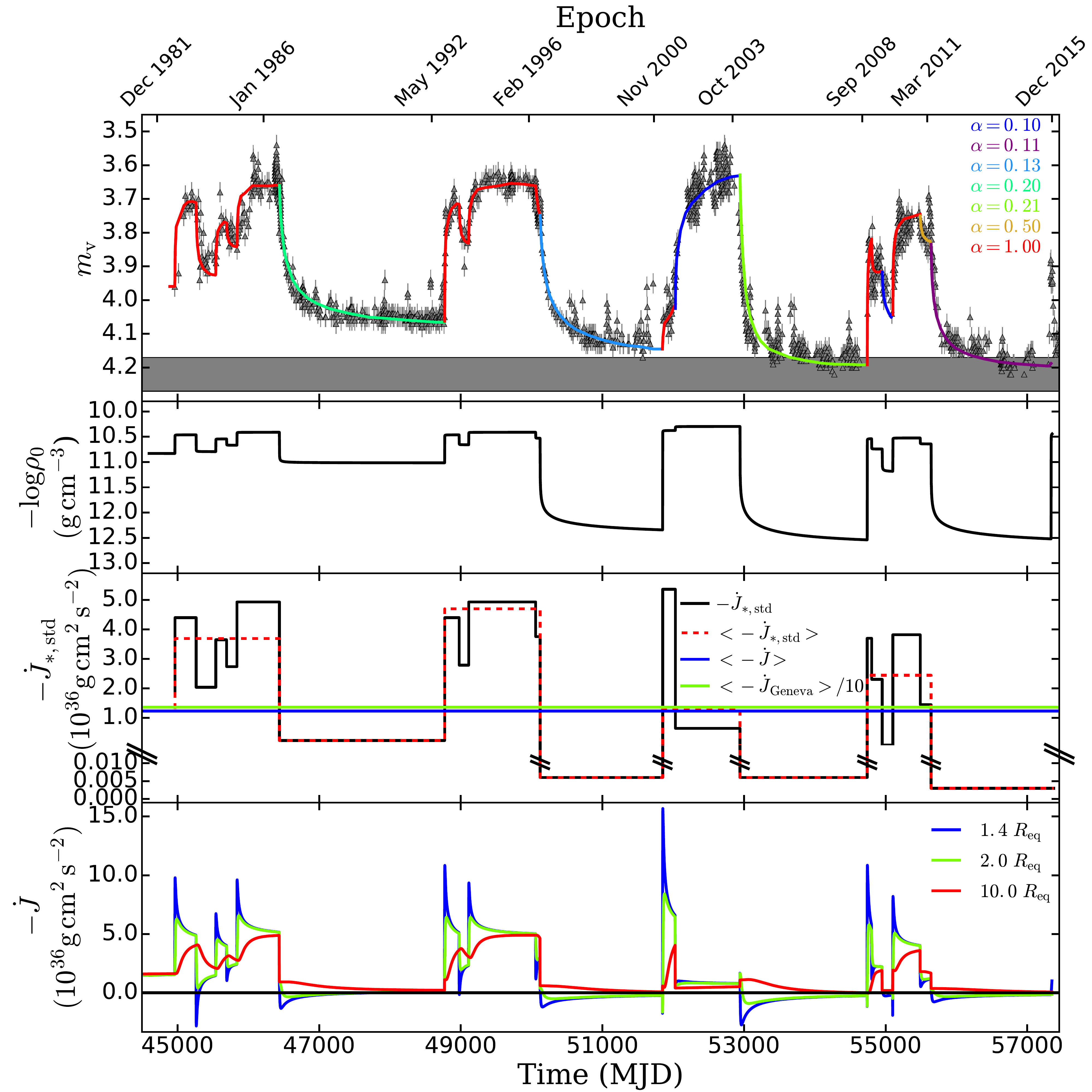}}
\caption{
First panel: Model fit of the full light curve of $\omega$ CMa. Each colored solid line represents an individual value for the $\alpha$ parameter, as indicated.
Second panel: Time evolution of the mass density at the base of the disc ($\rho_0$).
Third panel: The history of $-\dot{J}_{*,\mathrm{std}}$ of the best fit model (the black lines). The dashed red lines display the mean AM loss rate of the star at each phase. The blue line shows the mean AM loss rate of the star during 34 years. The green line shows ten percent of the mean AM loss during 34 years according to \citet{granada2013}. The ordinate is broken for better presentation of the low values against the high ones. 
Fourth panel: AM flux as a function of time, at different positions in the disc. The blue, green and red lines correspond to $r = 1.4\,R_\mathrm{eq}$, $r = 2.0\,R_\mathrm{eq}$, $r = 10.0\,R_\mathrm{eq}$, respectively.
}
\label{summary}
\end{figure*}

In Sect.~\ref{vdd_model} we presented the first physical model (VDD model) to fit the light curve of a Be star including several formation and dissipation phases. It was necessary to consider varying the $\alpha$ parameter to achieve a satisfactory result. The top panel of Fig.~\ref{summary} summarizes our model results. The light curve was fitted using higher values of $\alpha$ for formation phases while the dissipation phases needed lower values (see Sect.~\ref{temperature_evolution}). Moreover, we found that $\alpha$ is not related to the cycle to cycle variations, and is not increasing nor decreasing continuously. The separation between the horizontal grey band (expressing the intrinsic magnitude of the star) and the model curves suggests that $\omega$ CMa never experiences a true quiescence, but instead switches between a high-density phase (outburst) and a low-density one (dissipation). A true quiescence may only have been reached at the end of the last cycle.

The estimated values of $\alpha$ are in rough agreement with the values derived for dwarf novae \citep{king2007, kotko2012}, as well as with the values acquired for the SMC by \citet{rimulo2018}. They are, however, an order of magnitude or more above the usual values obtained in magnetohydrodynamic simulations that employ the 
magnetorotational instability \citep[MRI,][]{balbus1991} as a possible mechanism to explain the viscosity.

Some studies in the literature \citep[e.g.,][]{touhami2011, vieira2017} use the following power-law approximation for the disc density:
\begin{equation}
\label{volume_density}
\rho = \rho_\mathrm{0}\left(\frac{r}{R_\mathrm{eq}}\right)^{-n}\,,
\end{equation}
where $\rho_\mathrm{0}$ is the volume density at the inner rim of the disc (base density). The density slope $n$ varies between 1.5 -- 3.5 with a statistical peak probability around 2.4 \citep{vieira2017}. Studies of individual stars reported values between 1.5 to 4.2  for $n$ and $7 \times 10^{-13}$ to $4.5 \times 10^{-10}$ $\mathrm{g\,cm^{-3}}$ for $\rho_0$ (e.g., \citealt{carciofi2006b}, \citeyear{carciofi2007}, \citeyear{carciofi2009}, \citealt{gies2007}, \citealt{jones2008}, \citealt{tycner2008}, \citealt{klement2015}, and \citealt{vieira2015}). The values of $\rho_0$ reported in the literature can be compared to the ones obtained for $\omega$ CMa in the second panel of Fig.~\ref{summary}. We found that $\rho_{0}$ is varying between $2.9 \times 10^{-13}$ to $5.0 \times 10^{-11}\ \mathrm{g\,cm^{-3}}$ which is almost two orders of magnitude in range.


The life cycles of Be star discs were discussed by \cite{vieira2017}, who modeled the infrared emission of $80$ objects, at up to three different epochs (\textit{IRAS}, \textit{WISE} and \textit{AKARI} data). By adopting a simple power-law prescription for the disc density profile, they estimated the $\rho_0$ and $n$ values for this large sample. Their results suggest a strong correlation between these parameters that were interpreted in terms of the time-dependent VDD properties. For that purpose, they computed representative evolutionary tracks over the $n-\log\rho_0$ diagram, representing scenarios such as a long disc build-up followed by a full disc dissipation, and cyclic cases where build-up and dissipation alternate on a regular basis. Their results suggested an evolutionary interpretation of the disc as forming, steady-state or dissipating in different parts of the $n-\log\rho_0$ diagram.

\begin{figure}
\centering
\includegraphics[width=1.0\linewidth]{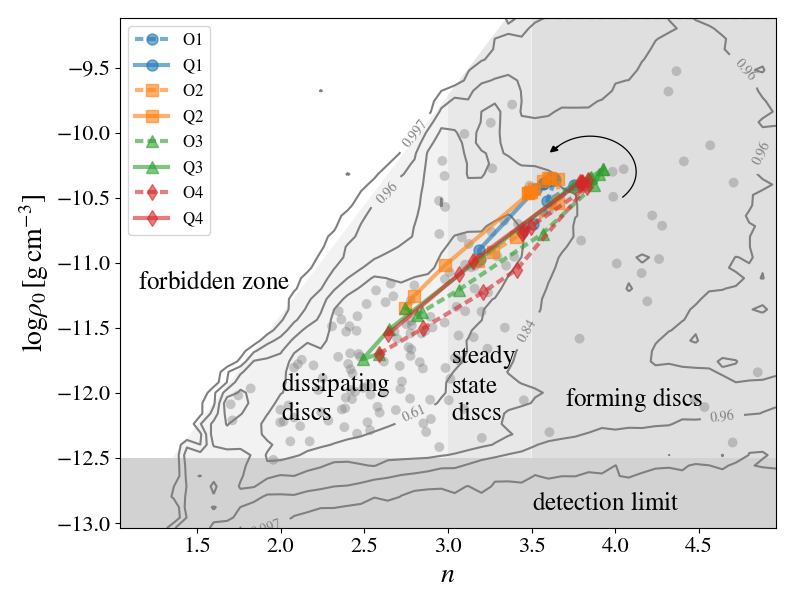}
\caption{Evolutionary tracks computed for each of $\omega$ CMa's cycles over the $n-\log\rho_0$ diagram. 
The epochs were selected to cover the beginning, middle and the end of each phase. These tracks were superimposed to the parameter distribution of the sample of \citet[][gray contours, integral normalized to unity]{vieira2017}. The evolutionary regions proposed by these same authors are also indicated (shaded regions). The scatter distribution of sample Galactic stars is represented by the grey circles.}
\label{evolution_track}
\end{figure}

Fig.~\ref{evolution_track} shows the tracks of $\omega$ CMa's photometric cycles over the $n-\log\rho_0$ diagram, calculated according to the procedure used by \cite{vieira2017}. Each  of $\omega$ CMa's cycles corresponds to a loop in this diagram. Major outbursts are accompanied by a rapid excursion to the forming discs region, where the disc has both a large base density and density slope. Once the outburst developed for a sufficiently long time, the track reaches the steady-state region, where the disc inner region is fully built-up (large $\rho_0$) and $n\simeq 3.5$. This situation is especially clear for the first two cycles, for which the build-up phases are longer.

Once the outburst ends, the mass injection rate drops abruptly. As a result, the base density decreases, and the density profile becomes flatter (smaller $n$). The subsequent loops grow wider with time, reaching smaller $n-\rho_0$ values for the last cycles. This happens because the quiescent level of $-\dot{J}_{\mathrm{*,std}}$ also decreases with time (Fig.~\ref{summary}), which causes the disc to reach smaller densities. 

\cite{vieira2017} concluded that there is a correlation between the spectral type and the disc density with the specific sub-population B2 type stars being more concentrated in the central region. The results for the case of $\omega$ CMa (B2 type star) shown in Fig.~\ref{evolution_track} are in very good agreement with those found for similar stars in the Galaxy. 


\subsection{Angular Momentum Loss}
\label{angular_momentum_loss}

\begin{figure}
\centering
\includegraphics[width=1.0\linewidth]{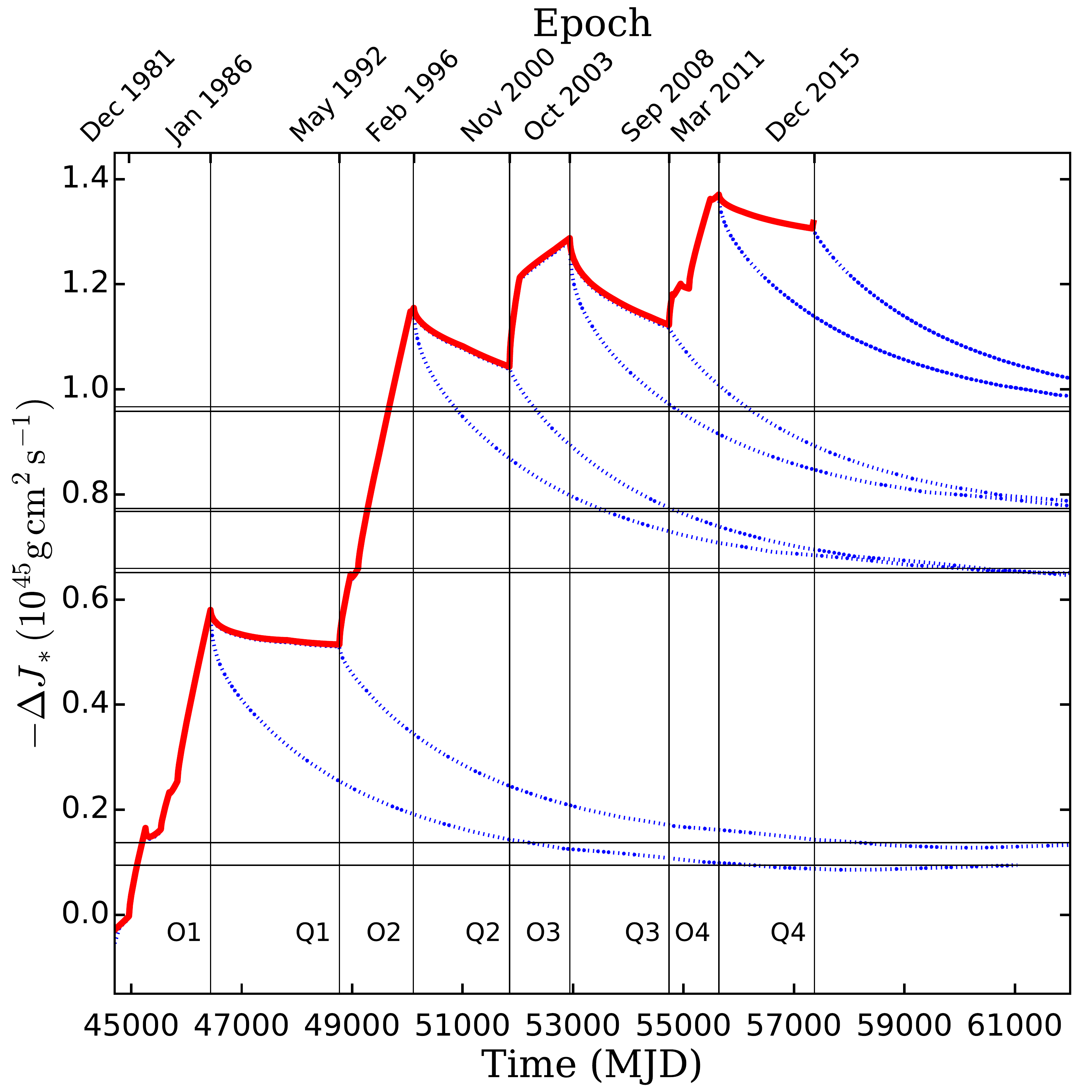}
\caption{
Thick solid red curve: AM lost by the star, $-\Delta J_*(t)$, from $\mathrm{MJD}_0=44960.5$ on. 
Dotted blue curves: Continuations of the AM lost by the star, if the injection of matter were to cease at that time with $\alpha=1$. 
The black horizontal lines are showing the integrated amount of AM lost at the end of each phase. 
}
\label{angular_momentum}
\end{figure}

The modeling of four complete cycles of $\omega$ CMa allows us to calculate the total AM lost by the star in the past 34 years. This quantity is given by
\begin{equation}
-\Delta J_*(t)=-\int_{t_0}^{t}\dot{J}_*(t')\mathrm{d}t'\,,
\label{deltaj}
\end{equation}
where $\dot{J}_*(t)$ comes from Eq.~(\ref{jdot_star}). In Fig.~\ref{angular_momentum}, the thick solid red curve shows $-\Delta J_*(t)$ for $t_0=44960.5$. 
The outburst phases O1, O2, O3 and O4 correspond to intense AM losses, while during quiescence (Q1, Q2, Q3 and Q4) some of the AM comes back to the star with the part of the disc matter that was reaccreted during these phases. At each inflection of the curve, representing the transition between outburst and quiescence or vice-versa, the dotted blue curves show the AM that would have been lost by the star had the AM injection been completely ceased from that instant on. The horizontal lines indicate the asymptotic value of these curves. Because in $\omega$ CMa the quiescence phases are not true quiescences, a net AM is still lost in these phases. 

\begin{figure}
\centering
\includegraphics[width=1.0\linewidth]{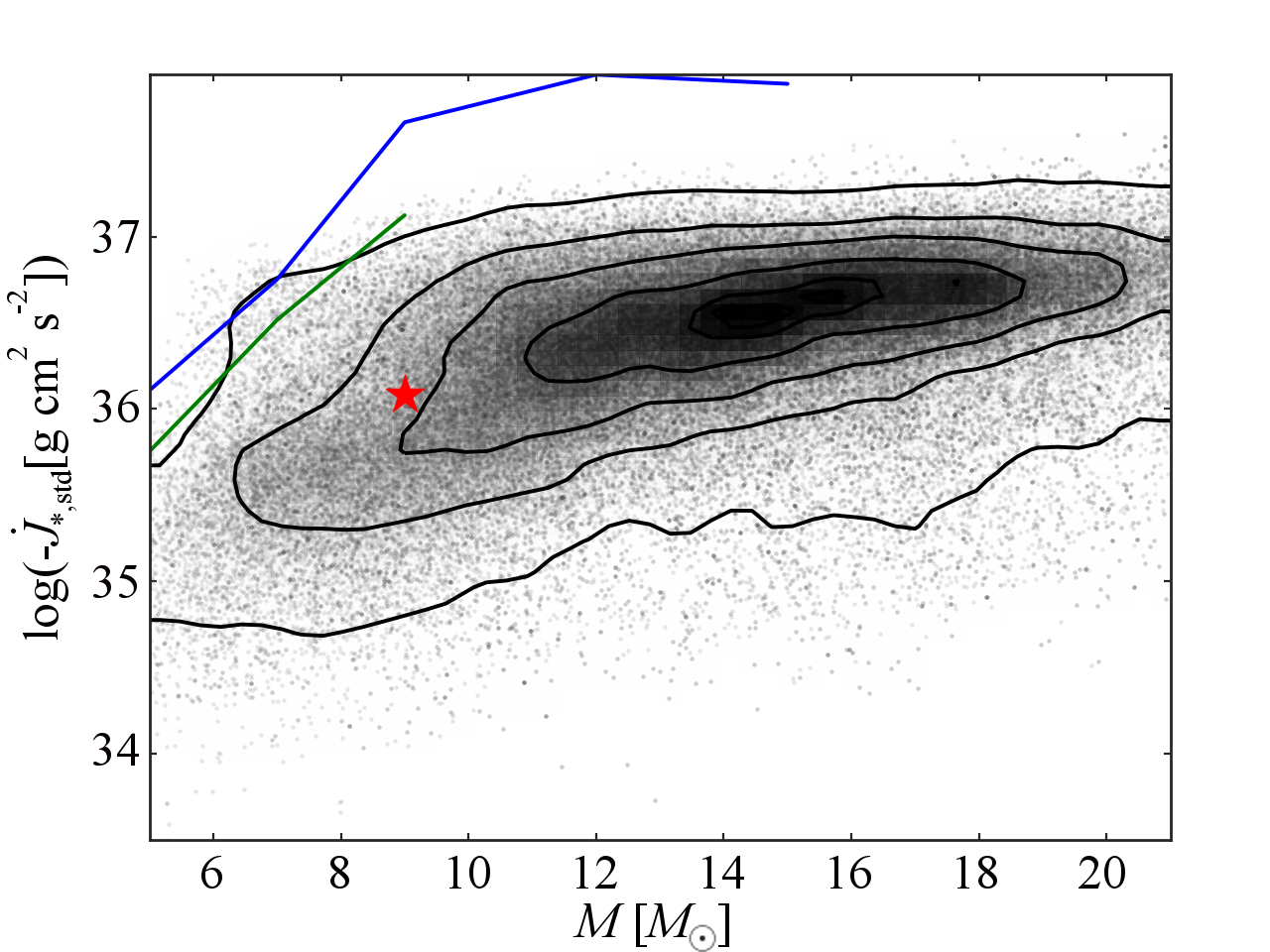}
\caption{The position of $\omega$ CMa in $-\dot{J}_{*,\mathrm{std}}$ vs $M$ diagram calculated by \citet{rimulo2018} for the Be stars in the SMC, in comparison to the estimated values by \citet{granada2013} for the Galaxy (green curve) and the SMC (blue curve).}
\label{logjdot}
\end{figure}

It has been proposed (e.g., \citealt{krticka2011}) that, as the star evolves through the main sequence, the formation of the VDD would be a natural mechanism to extract AM from the outer layers of the star, preventing it from exceeding the break up velocity. \cite{granada2013}, using the Geneva stellar evolution code, 
estimated the theoretical rate of AM loss expected during the main sequence evolution of stars with different masses and metallicities. To do so, they assumed the appearance of a steady-state viscous decretion disc  every time the outer layers of the star reached a given rotation rate limit. \cite{rimulo2018} found that the predictions of the Geneva code for AM loss rates are much larger than the actual rates observed in Be stars from the SMC (compare the blue line with the shaded areas in Fig.~\ref{logjdot}). Our results suggest a similar disagreement for the Galactic Be star $\omega$ CMa. For a star of Galactic metallicity and 9.0\,M$_{\odot}$, the AM loss rate computed by \cite{granada2013} was $1.3\times 10^{37}\,\mathrm{g\,cm^2\,s^{-2}}$ (green line in the third panel of Fig.~\ref{summary}), which is almost eleven times larger than the mean value estimated from this work. Incidentally, it is interesting to point out that $\omega$ CMa loses AM at a rate similar to that of a similar 9.0\,M$_{\odot}$ Be star in the SMC. Therefore, at this point our result does not indicate any difference in the AM loss rate as a function of metallicity. Clearly, to properly address this issue a large sample of Galactic Be stars must be investigated.

The third panel of Fig.~\ref{summary} shows the evolution of $-\dot{J}_{*,\mathrm{std}}$ during 34 years (black line) that varies between $3.0 \times 10^{33}$ to $5.4 \times 10^{36} \mathrm{g\,cm^2\,s^{-2}}$. During the formation phases the value of $-\dot{J}_{*,\mathrm{std}}$ increased noticeably, except for the O3 section that corresponds to a low value of the $\alpha$ parameter. The dissipation phases match the decrease in $-\dot{J}_{*,\mathrm{std}}$ but may never go to zero. The red lines represent the mean AM loss rates of the star in each phase. The blue horizontal line represents the mean AM loss rate in the period of 34 years ($<-\dot{J}>$) covered by the light curve of $\omega$ CMa. This rate is $1.2\times 10^{36}\,\mathrm{g\,cm^2\,s^{-2}}$, and the total AM lost in the whole studied period was $1.3\times 10^{45}\,\mathrm{g\,cm^2\,s^{-1}}$. From the Geneva evolutionary models, a B2 star with W=0.73 has a total AM of $2.1\times 10^{53}\,\mathrm{g\,cm^2\,s^{-1}}$. Thus, the AM lost over 34 years, is just $6\times 10^{-9}$ of that. Even if we extrapolate the measure of AM loss rate for the entire main sequence lifetime, which is about $3.25\times 10^7$ yr, this would correspond to 0.006 of the total AM of the star. 

The fourth panel of Fig.~\ref{summary} demonstrates the variation of the AM flux ($-\dot{J}$) for the four cycles of $\omega$ CMa at different radii (1.4, 2.0 and $10.0\,R_\mathrm{eq}$). While $-\dot{J}$ has both positive (losing) and negative (gaining) values at 1.4 and $2.0\,R_\mathrm{eq}$ it never has a negative value at the outer radius $10.0\,R_\mathrm{eq}$ meaning that, from this distance outward, the disc always remains a decretion disc \citep{rivinius2013b}.

\cite{rivinius1999} provided strong evidence for the quasi-Keplerian nature of Be discs even before they were resolved by interferometry. The values of AM presented above correspond to the radial velocity ($v_{r}$) between $\sim$ -3 to $\sim$4 km/s in the $\omega$ CMa's disc which is in good agreement with the typical values presented in \cite{rivinius1999}. Moreover, comparing these values to the orbital velocity ($v_\mathrm{orb}$) of $\sim$500 km/s in the inner rim of the disc, we find that our results match the detailed hydrodynamic simulations which show that for most of the disc (within at least several tens of stellar radii), the azimuthal velocity is much larger than the radial velocity \citep{krticka2011}.


\subsection{Evolution of the disc temperature}
\label{temperature_evolution}


\begin{figure*}
\centering
\includegraphics[width=1.0\linewidth]{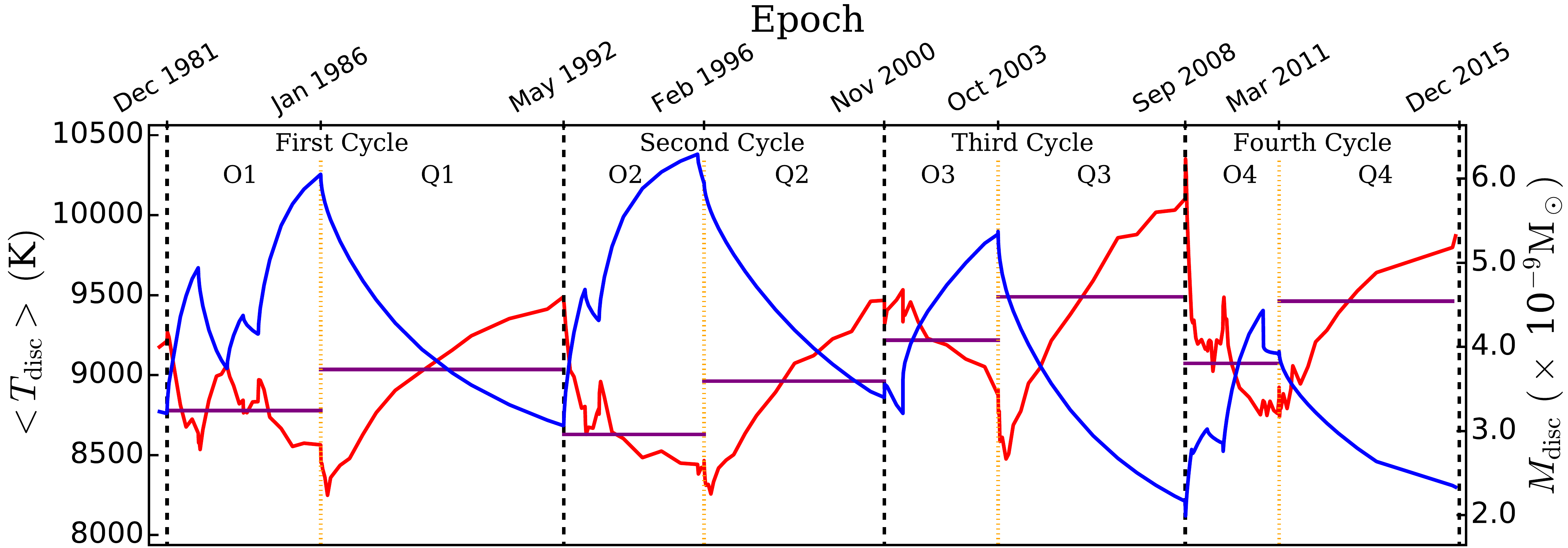}
\par\medskip
\centering
{\includegraphics[width=1.0\linewidth]{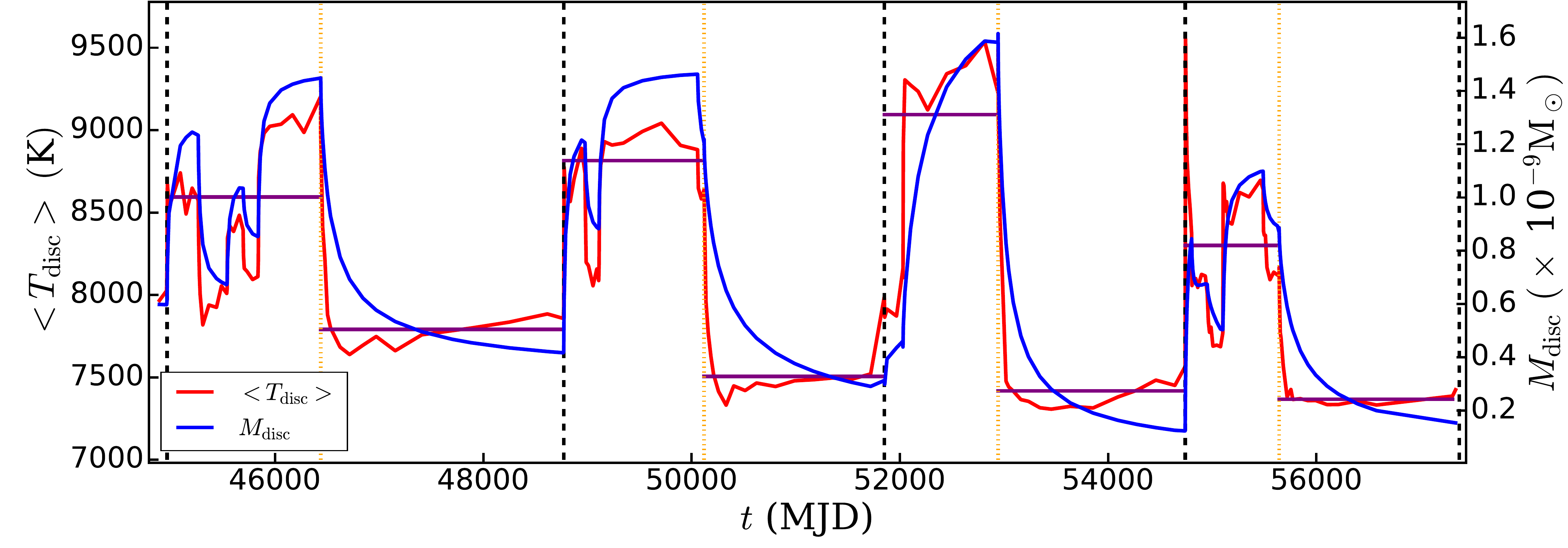}}
\caption{Evolution of mass-averaged temperature (red line, Eq.~\ref{mean_disc_temp}) and total mass (blue line) of $\omega$ CMa's disc for the whole (top panel) and inner part ($r < 3.0\,R_{\rm eq}$) of the disc (bottom panel). The horizontal purple lines represent the mean temperature during each phase.}
\label{disc_temperature}
\end{figure*}


\begin{figure}
\centering
\includegraphics[width=1.0\linewidth]{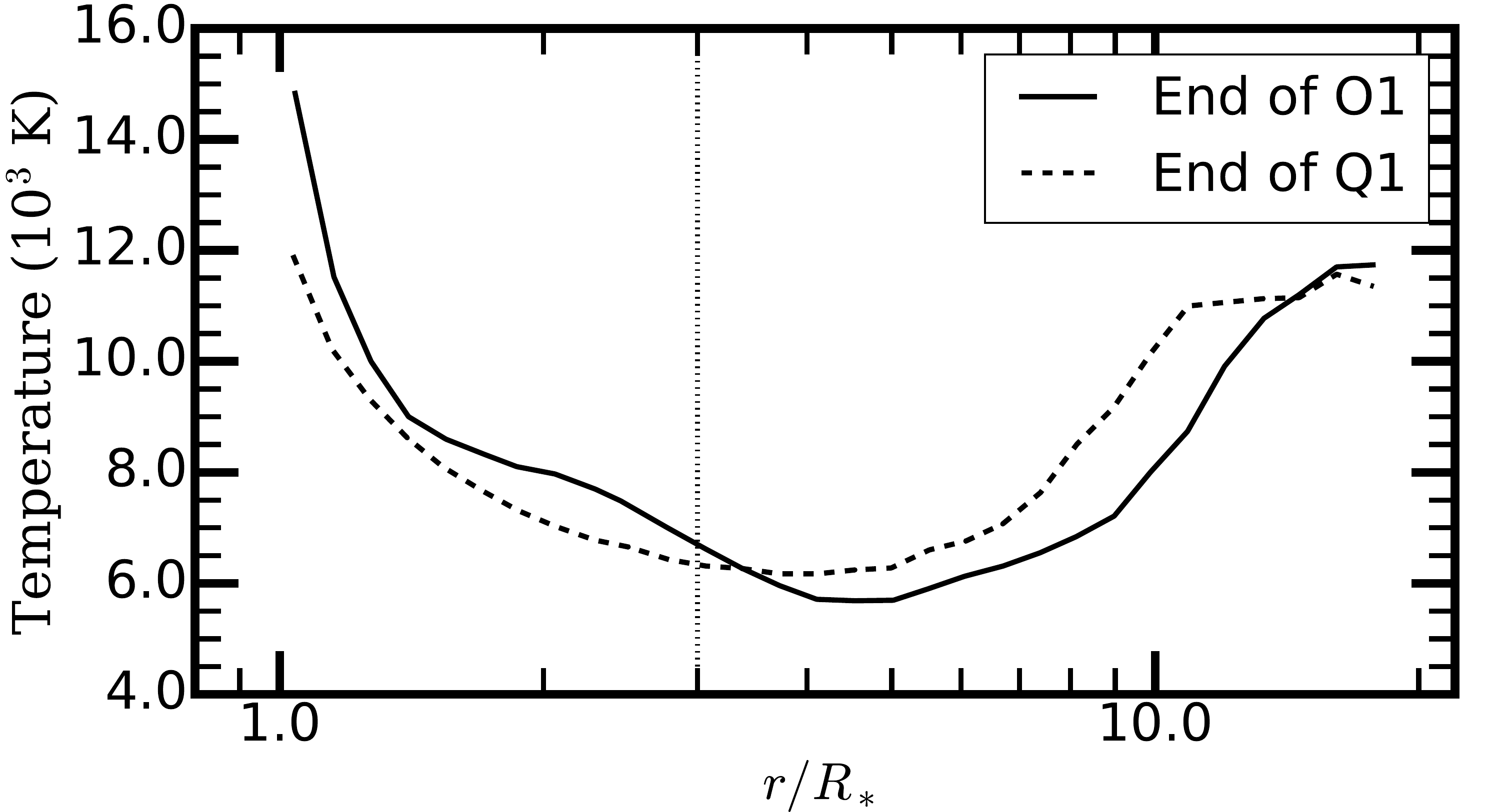}
\caption{Temperature of the disc as a function of radius at the end of O1 (solid line) and Q1 (dashed line) representing the fact that the inner disc is hotter in the high-mass state while the bulk of the disc is hotter in the low-mass state.}
\label{temperature_vs_r}
\end{figure}


One of the significant results of this work is that, with the exception of O3, the values of $\alpha$ are larger during the formation phase than during dissipaiton. As described in Sect.~\ref{vdd_c3}, however, the result of O3 may be of little statistical significance as the early stages of the outburst are poorly sampled. This result is consistent with the findings of \cite{rimulo2018} who studied 81 disc formation and dissipation phases of 54 Be stars in the Small Magellanic Cloud. As Eq.~(\ref{sigmadot}) shows, the timescale of disc evolution is set by the factor $\alpha T$. In the {\tt SINGLEBE} hydrodynamic calculations the disc was set to a constant isothermal temperature of $T_\mathrm{isot}=13200\,\mathrm{K}$, following \cite{carciofi2006a}. However, the {\tt HDUST} simulations that are run for selected times of the disc evolution calculate the coupled problem of radiative and statistical equilibrium, thus determining the gas kinematical temperature as a function of position and time. 
In this section we investigate whether this inconsistency in the modeling might be responsible for $\alpha$ being larger at build-up than dissipation.

We begin by defining the mass-averaged temperature of the disc inside a spherical shell of inner radius $R_\mathrm{eq}$ and outer radius $r$
\begin{equation}
\left\langle T_\mathrm{disc}\right\rangle_r(t)=\frac{1}{M_{\mathrm{disc},r}}\int_{R_\mathrm{eq}}^r\oint_{4\pi} T_\mathrm{disc}(\boldsymbol{s},\Omega,t)\rho(\boldsymbol{s},\Omega,t) \mathrm{d}\Omega s^2\mathrm{d}s\,,
\label{mean_disc_temp}
\end{equation}
where $M_{\mathrm{disc},r}$ is the mass of the disc inside the same spherical shell, given by 
\begin{equation}
M_{\mathrm{disc},r}(t)=\int_{R_\mathrm{eq}}^r\oint_{4\pi}\rho(\boldsymbol{s},\Omega,t) \mathrm{d}\Omega s^2\mathrm{d}s\,,
\label{mass_disc_r}
\end{equation}
and $T_\mathrm{disc}$ is the disc temperature as a function of position as calculated by {\tt HDUST}.

Fig.~\ref{disc_temperature} shows the evolution of $\left\langle T_\mathrm{disc}\right\rangle_r$ and $M_{\mathrm{disc},r}$ for two values of $r$: $r=3R_\mathrm{eq}$, corresponding to the inner disc, where most of the visual continuum flux comes from (lower panel), and $r=18R_\mathrm{eq}$, corresponding to the whole simulated disc by {\tt HDUST} (upper panel).

The behavior of $M_{\mathrm{disc},3}(t)$ and $M_{\mathrm{disc},18}(t)$ qualitatively follows the behavior of the light curve, as expected, since the formation and dissipation phases are associated with increasing and decreasing disc mass. 
The rate of evolution of $M_{\mathrm{disc},3}(t)$ is much faster than $M_{\mathrm{disc},18}(t)$ because the viscous timescale scales as $r^{0.5}$. In other words, the inner disc reacts much faster to changes in the disc feeding rate than the whole disc does. The mass-averaged temperature, however, presents a different behavior: in the inner disc, the average temperature is positively correlated with the disc mass, but the opposite happens for the whole disc. To understand this, we plotted the temperature profile of $\omega$ CMa at the end of O1 (high-mass state) and at the end of Q1 (low-mass state) in Fig.~\ref{temperature_vs_r}, that shows that the inner disc is hotter in the high-mass state while the bulk of the disc is hotter in the low-mass state which naturally explains the above correlations.

From Fig.~\ref{disc_temperature} we conclude that the average temperature at the inner disc during build-up is larger than during dissipation by $\sim 800$--$1700\,\rm K$, and the reverse is true for the whole disc, but with smaller temperature differences ($\sim 250$--$350\,\rm K$). There is spectroscopic evidence for such variations for the stars $\delta$ Sco and $\eta$ Cen, notably the coexistence/balance between FeIII and FeII emission/absorption, but also HeI/HeII (unpublished data).

Let us now examine the effects that the temperature differences in both phases might have on the $\alpha$ determination. Any combination of $\alpha$ and $T$ that keeps $\alpha T$ unchanged does not affect the surface density solution $\Sigma(r,t)$. Therefore, assuming that the \emph{real} value of the viscosity, $\alpha_{\rm real}$, is associated with the temperature calculations shown in Fig.~\ref{disc_temperature}, by using an isothermal disc temperature in the hydrodynamic calculations we introduce a bias in the $\alpha$ determination given roughly by
\begin{equation}
\frac{\alpha_{\rm biased}}{\alpha_{\rm real}} = \frac{\left\langle T_\mathrm{disc}\right\rangle_r}{T_\mathrm{isot}}\,.
\label{mean_disc_temp2}
\end{equation}
Here, we ignore possible temporal and radial variations of $\alpha$.
Given that the hydrodynamic calculations are isothermal, this bias is certainly present in our $\alpha$ determinations. However, the effect is likely small, as the temperature variations between the different phases amount to only $~5$--$10\%$ (Fig.~\ref{disc_temperature}), which means that the bias in $\alpha$ would be at most around 10\%. To properly address this issue, non-isothermal hydrodynamic calculations, coupled with radiative transfer, should be performed, which is beyond the scope of this paper. However,  because the calculated differences in $\alpha$ for the build-up and dissipation phases are much larger than the above, we conclude that these differences are real. 

On the other hand, in a recent paper, \cite{kee2016} discuss the role of radiative ablation in the mass budget of Be discs. They show that radiation forces can play an important role in dissipating the disc, and some of their model calculations show that a disc wind can ablate the entire disc in timescales of order months to years comparable to what is observed. However, their model calculations were performed in the optically thin approximation, which clearly do not apply in our case. 
Therefore, it is possible that radiative ablation plays a role in Be disc dynamics, but the extent of this role remains to be determined. In principle, if radiative ablation is important in $\omega$ CMa's case, the values of $\alpha$ quoted in this paper are likely upper limits. Future work is necessary to answer this question.


\section {Conclusions}
\label{conclusions}

We investigated the $V$-band light curve of $\omega$ CMa that, in the past 34 years, underwent four complete cycles of disc formation followed by a partial dissipation. Typically, formation phases lasted between $\sim$ 2.5--4.0 years and the dissipation phases $\sim$ 4.5--6.5 years. The results of a detailed VDD hydrodynamic modeling coupled with 3-D radiative transfer calculations suggest six main conclusions:

\begin{itemize}

\item[1)] We demonstrate that the VDD model is capable of reproducing the disc variability during both build-up and dissipation phases. 
\item[2)] Different values of $\alpha$, in the range of $0.1-1.0$, were required to model the data. Typically, the values of $\alpha$ during the formation phases are higher than during dissipation. A similar trend was recently found for a sample of Be stars in the SMC \citep{rimulo2018}. The reader is referred to this work for a discussion of possible causes for this phenomenon.
\item[3)] Even during apparent quiescence, the disc feeding by the star does not seem to fully vanish. On time scales of a month or two, the star always ejects matter.  
\item[4)] Depending on the distance from the star, the AM flux may be positive (decretion) or negative (accretion). The average AM lost by the star is always positive, even during apparent quiescence.
\item[5)] The total AM lost by the star during the 34 years of observations is $1.3\times 10^{45}\,\mathrm{g\,cm^2\,s^{-1}}$ which is $6\times 10^{-9}$ of the total AM of the central star. If extrapolated for the whole main sequence, this would suggest that about 0.006 of the initial AM content of the star would be lost. This value is about eleven times smaller than the requirements of the Geneva evolutionary models of fast rotating stars. This result agrees with the study of 54 Be stars in the SMC done by \cite{rimulo2018}.
\item[6)] When compared to Galactic Be stars of similar spectral type, $\omega$ CMa displays a similar disc density scale.

In a forthcoming study, we will extend the analysis made in this paper to other observables, namely emission line profiles, continuum polarization and IR spectrointerferometry. As these observables originate in different parts of the disc, this analysis will allow us to tackle the interesting question of whether the viscosity parameter varies with the distance from the star or not.

\end{itemize}


\section*{acknowledgements}
This work made use of the computing facilities of the Laboratory of Astroinformatics (IAG/USP, NAT/Unicsul), whose purchase was made possible by the Brazilian agency FAPESP (grant 2009/54006-4) and the INCT-A. Also, this work has been supported financially by the Research Institute for Astronomy \& Astrophysics of Maragha (RIAAM) under research project No. 1/5237-63. M.R.G. acknowledges the support from CAPES PROEX Programa Astronomia. A.C.C. acknowledges support from CNPq (grant 307594/2015- 7) and FAPESP (grant 2015/17967-7). L.R.R. acknowledges the support from FAPESP (grant 2012/21518-5) and from CNPq (grant 142411/2011-6). R.G.V. acknowledges support from FAPESP (grant No 2012/20364-4). D.M.F acknowledges support from FAPESP (grant 2016/16844-1). J.E.B acknowledges support from NSF grant AST-1412135.



\bibliographystyle{mnras}
\bibliography{references} 



%
%


\bsp	
\label{lastpage}
\end{document}